\documentclass[%
 reprint,
superscriptaddress,
 amsmath,amssymb,
 aps,
]{revtex4-2}

\usepackage{graphicx}
\usepackage{dcolumn}
\usepackage{bm}
\usepackage{hyperref}
\hypersetup{
	colorlinks=true,
	linkcolor=nrppurple,
	citecolor=nrppurple,
	urlcolor=nrppurple,
}

\newcommand{\del}{\partial}
\renewcommand{\v}[1]{\bm{\mathrm{#1}}}

\renewcommand{\k}{\v{k}}

\newcommand{\df}{\mathrm{d}}
\newcommand{\Df}{\mathcal{D}}
\newcommand{\tr}{\mathrm{tr}}

\newcommand{\pfrac}[2]{\left( \frac{#1}{#2} \right)}

\newcommand{\goes}{\rightarrow}

\newcommand{\sqm}[4]{\begin{pmatrix}#1 & #2 \\ #3 & #4 \end{pmatrix}}
\newcommand{\twovec}[2]{\begin{pmatrix}#1 \\ #2 \end{pmatrix}}

\newcommand{\tworow}[2]{\begin{pmatrix}#1 & #2 \end{pmatrix}}

\newcommand{\e}{\mathrm{e}}
\newcommand{\w}{\omega}
\newcommand{\W}{\Omega}

\makeatletter \renewcommand\@make@capt@title[2]{%
\@ifx@empty\float@link{\@firstofone}{\expandafter\href\expandafter{\float@link}}%
\sffamily{\textbf{#1}}\@caption@fignum@sep#2 }
\usepackage{placeins} 

\begin{document}
\definecolor{nrppurple}{RGB}{128,0,128}

\preprint{APS/123-QED}

\title{Spectroscopic signatures of time-reversal symmetry breaking superconductivity}

\author{Nicholas R. Poniatowski}
\email[]{nponiatowski@g.harvard.edu}
\affiliation{Department of Physics, Harvard University, Cambridge, MA 02138}

\author{Jonathan B. Curtis}
\affiliation{Department of Physics, Harvard University, Cambridge, MA 02138}
\affiliation{John A. Paulson School of Engineering and Applied Sciences, Harvard University, Cambridge, MA 02138}

\author{Amir Yacoby}
\affiliation{Department of Physics, Harvard University, Cambridge, MA 02138}
\affiliation{John A. Paulson School of Engineering and Applied Sciences, Harvard University, Cambridge, MA 02138}

\author{Prineha Narang}
\email[]{prineha@seas.harvard.edu}
\affiliation{John A. Paulson School of Engineering and Applied Sciences, Harvard University, Cambridge, MA 02138}

\date{\today}

\begin{abstract}
The collective mode spectrum of a symmetry-breaking state, such as a superconductor, provides crucial insight into the nature of the order parameter. In this context, we present a microscopic weak-coupling theory for the collective modes of a generic multi-component time-reversal symmetry breaking superconductor, and show that fluctuations in the relative amplitude and phase of the two order parameter components are well-defined underdamped collective modes, even in the presence of nodal quasiparticles. We then demonstrate that these ``generalized clapping modes'' can be detected using a number of experimental techniques including ac electronic compressibility measurements, electron energy loss spectroscopy, microwave spectroscopy, and ultrafast THz spectroscopy.
Finally, we discuss the implications of our work as a new form of ``collective mode spectroscopy'' that drastically expands the number of experimental probes capable of detecting time-reversal symmetry breaking in unconventional superconductors such as Sr$_{\text{2}}$RuO$_{\text{4}}$, UTe$_{\text{2}}$, and moir\'e heterostructures. 
\end{abstract}

\maketitle

There is a rich and constantly expanding taxonomy of unconventional superconducting phases, and increasingly sophisticated probes are needed to distinguish one such phase from another. A defining feature of any superconducting state is its collective mode spectrum, which encodes the dynamics of the order parameter. The collective modes of conventional (phonon-mediated) superconductors are well-established, consisting of two modes which correspond to fluctuations in the amplitude and phase of the order parameter. The first, called the Higgs mode \cite{anderson-58-2,Schmid1968}, is massive and resides at the edge of the quasiparticle continuum \cite{higgs-jetp,Kulik1981}, while the second, called the Anderson-Bogoliubov-Goldstone (ABG) mode \cite{Bogo-original,anderson-58-1}, is massless in a neutral system, in accordance with Goldstone's theorem, but is lifted to the plasma frequency in the presence of long-ranged Coulomb interactions \cite{anderson-63}, and is thus indistinguishable from the usual plasmon in real materials.

However, systems with more complex order parameters can exhibit a rich collective mode spectrum featuring other modes, such as additional Higgs modes in anisotropically gapped (e.g. $d$-wave) superconductors \cite{varma-dwave}, or Leggett modes in multiband superconductors \cite{leggett-mode}. The presence of these additional modes in the spectrum can then be taken as a fingerprint of the underlying order parameter symmetry. That is, the dynamics of the order parameter can be studied to gain insight into its equilibrium structure. Such a scheme has only recently been proposed in the context of anisotropically gapped superconductors, where the spectrum of non-equilibrium Higgs modes can be used to deduce the orbital symmetry of the order parameter \cite{non-eq-higgs-expt,higgs-spect-th}.  

In this article, we generalize this notion of ``collective mode spectroscopy'' to a particularly exotic class of unconventional superconductors, namely those which spontaneously break time-reversal symmetry in addition to global $U(1)$ symmetry at the superconducting transition. These time-reversal symmetry breaking (TRSB) superconducting states are the subject of considerable current interest, and are believed to be realized in a number of bulk materials, including Sr$_2$RuO$_4$ \cite{214-kerr, 214-musr}, UPt$_3$ \cite{upt3-kerr,upt3-musr}, URu$_2$Si$_2$ \cite{uru2si2-kerr}, UTe$_2$ \cite{ute2-original,ian-ute2}, PrOs$_4$Sb$_{12}$ \cite{prossb-kerr,pr-musr}, and K-doped BaFe$_2$As$_2$ \cite{pnictide}, as well as engineered structures such as Bi/Ni bilayers \cite{bini-kerr} and SnTe nanowires \cite{jimmy}. Moreover, recent theoretical proposals have suggested that such states may also be realized in moir\'e heterostructures \cite{magic-angle-d+id,magic-angle-d+id-2}, and could be generically engineered in twisted bilayers of anisotropically gapped superconductors \cite{twisted-bscco, volkov2020magic}. 

In what follows, we identify two collective modes, the ``generalized clapping modes,'' which are unique to TRSB superconductors and subsequently derive their spectrum from a generally applicable microscopic weak-coupling theory. Our first key finding is that these modes are coherent collective excitations for \emph{generic} TRSB states, even those which have point or line nodes in the superconducting gap. Owing to the universality of these modes in TRSB superconductors, they may serve as clear spectroscopic signatures which identify a TRSB superconducting state. To this end, we discuss a variety of existing experimental probes that can couple to these modes, and hence can be used as new means to detect TRSB superconductivity in quantum materials. We pay special attention to the superconducting state of Sr$_2$RuO$_4$, where the generalized clapping mode spectrum can distinguish the two current leading candidate order parameters. 

\textbf{\textsf{Generalized clapping modes.}} 
TRSB superconducting states are characterized by a doubly-degenerate complex multi-component order parameter of the form $\Delta = \Delta_1 \pm i\Delta_2$ (see appendix A), and can be divided into two classes: {\it (1)} systems where both components $\Delta_1$ and $\Delta_2$ belong to the same multi-dimensional irreducible representation (irrep) of the crystalline point group, in which case $|\Delta_1| = |\Delta_2|$ is required by symmetry, and {\it(2)} ``mixed symmetry'' systems where the two components belong to different irrep's, which can arise due to either an accidental degeneracy between two pairing channels or two successive superconducting transitions (see appendix A), in which case the ratio $|\Delta_1|/|\Delta_2|$ is unconstrained. 

Given this internal orbital structure of the order parameter, we would expect that in addition to the usual ABG and Higgs modes, there should be a massive mode corresponding to fluctuations of the relative phase between the two order parameter components around its equilibrium value of $\pm \pi/2$. Further, we expect that there should be a second amplitude mode which corresponds to fluctuations of the relative amplitude $|\Delta_1|/|\Delta_2|$ so there are a total of four real modes.

Historically, similar modes corresponding to ``internal vibrations of the structure of the order parameter'' were first recognized in $^\text{3}$He-A \cite{wolfe-cm}, one of which was named the ``clapping mode.'' This was subsequently extended to the two-dimensional chiral $p$-wave ($p+ip$) superconductor \cite{214-clapping}, where there are two such clapping modes which are degenerate and reside at a frequency $\W = \sqrt{2} \Delta_0$, with $\Delta_0$ the magnitude of the order parameter. 

As argued above, we can anticipate analogous modes, which we call the  generalized clapping modes, for any multi-component TRSB state on symmetry grounds alone \cite{balatsky-prl}. Although it is intuitively obvious that the generalized modes should exist in principle, there is no reason to expect \textit{a priori} that they are not overdamped by quasiparticle excitations. For a fully gapped superconductor, this requires that the frequencies of both modes lie below the quasiparticle continuum or, in the case of a nodal quasiparticle gap function, requires that the spectral function of each mode retains sharp features despite the presence of nodal quasiparticles. To establish that the generalized clapping modes are coherent, and thus experimentally detectable, for a generic TRSB state requires a derivation of their spectrum starting from a microscopic theory, which we furnish below. By studying the generalized clapping mode spectrum for a wide variety of TRSB order parameters we are able to establish the general features of these modes, which ultimately enable their use in the collective mode spectroscopy of real materials.

\begin{figure}
    \centering
    \includegraphics[width=70mm]{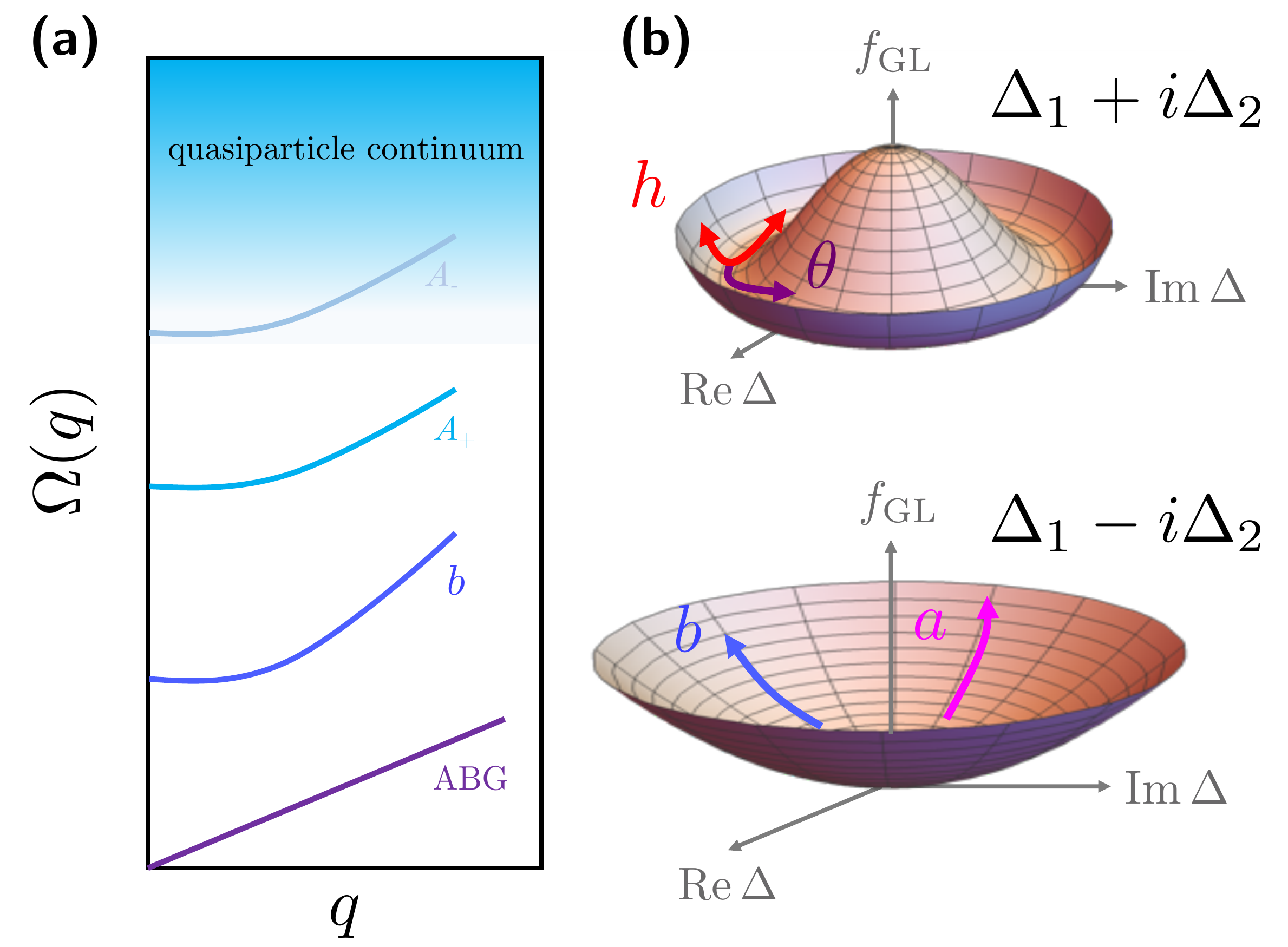}
    \caption{\textbf{Generalized clapping modes. }\textbf{(a)} Schematic of the collective mode spectrum for a TRSB superconductor, featuring the massless ABG mode (acoustic plasmon), massive relative phase mode (optical plasmon), and two massive amplitude modes. The $A_+$ mode generically lies below the quasiparticle continuum. \textbf{(b)} Illustration of the collective modes: once the system condenses into e.g. the $\Delta_1 + i\Delta_2$ ground state, the Higgs (ABG) mode is an amplitude (phase) fluctuation in the condensed pairing channel, while the generalized clapping modes can be thought of as amplitude and phase fluctuations in the time-reversed uncondensed pairing channel.}
    \label{schematic-fig}
\end{figure}

\begin{table*}[t]
\centering
\begin{tabular}{|c|c|c|c|c|l|}
\hline
     \textbf{Name} & \textbf{Irrep} & \textbf{Basis function} & \textbf{Type} & \textbf{Dimension} & \textbf{Candidate materials} \\
     \hline 
     $p+ip$ & $E_u$ & $k_x + ik_y$ & multi-dim & 2d & ??? \\
     $d+id'$ & $B_{1g} + iB_{2g}$ & $k_x^2 - k_y^2 + ik_x k_y$ & mixed symmetry & 2d & cuprates; moir\'e heterostructures \\
     $s+id$ & $A_{1g} + iB_{2g}$ & $1+ik_x k_y$ & mixed symmetry & 2d & pnictides \\
     $d+ig$ & $B_{1g} + iA_{2g}$ & $(k_x^2 - k_y^2) (1 + ik_x k_y)$ & mixed symmetry & 2d & Sr$_{\text{2}}$RuO$_{\text{4}}$ \\
     \hline 
     ABM & $E_u$ & $k_x + ik_y$ & multi-dim & 3d & $^{\text{3}}$He-A; possible caricature of UTe$_{\text{2}}$ \\
    $d+id$ & $E_g$ & $(k_x + ik_y)k_z$ & multi-dim & 3d & Sr$_{\text{2}}$RuO$_{\text{4}}$, URu$_{\text{2}}$Si$_{\text{2}}$  \\
    3d $d+id'$ & $B_{1g} + iB_{2g}$ & $k_x^2 - k_y^2 + ik_x k_y$ & mixed symmetry & 3d & SrPtAs, URu$_{\text{2}}$Si$_{\text{2}}$ \\
    \hline
\end{tabular}
\caption{ \textbf{TRSB order parameters.} For each order parameter we consider in this work, we list the corresponding irreducible representation (irrep) of the tetragonal point group, a representative basis function, the type of TRSB state, i.e. whether it belongs to a multi-dimensional irrep or is a ``mixed symmetry'' state (see appendix A), the spatial dimension of the Fermi surface it exists on, and candidate materials where such a state is believed to be realized. }
\label{table}
\end{table*}

\textbf{\textsf{Weak-coupling theory.}} We begin with a single band of fermions subject to the attractive interaction $V_{\k\k'} = -\sum_{\ell=1,2} g_{\ell} \chi^\ell_{\k}\chi^{\ell}_{\k'}$  where $g_{\ell} > 0$ are coupling constants and $\chi_{\k}^\ell$ are form factors which encode the orbital symmetry of the interaction. We take these to be real and normalized according to the inner product $\int \frac{\df \phi_{\k}}{2\pi} \, \chi^\ell_{\k} \chi^{\ell'}_{\k} = \delta_{\ell \ell'}$. We assume pairing in the $S^z_{\rm tot} = 0$ sector, but within this sector our results are applicable to both singlet and $m=0$ triplet pairing. We treat this system within the imaginary-time path-integral formalism by introducing a Hubbard-Stratonovich decoupling field $\Delta^{\ell}$ in each pairing channel and integrating out the fermions.
One then arrives at the effective action for the order parameters $\Delta^\ell $ of 
\begin{equation} \label{generic-sc-act}
    S = \sum_{q} \left( g_1^{-1} |\Delta^{(1)}_q|^2 + g_2^{-1} |\Delta^{(2)}_q|^2 \right) - \tr \log \mathbb{G}^{-1}.
\end{equation}
The inverse fermion propagator is $\mathbb{G}^{-1}_{k+q,k} = (i\w_n - \xi_{\k} \, \tau_z) \delta_{q,0} + \sum_{\ell} \Delta_q^{\ell} \chi_{\k}^{\ell} \, \tau^+ + \sum_{\ell} \bar{\Delta}_{-q}^{\ell} \chi^{\ell}_{\k} \, \tau^- $ where $\tau_i$ are the Pauli matrices in Nambu space, $\tau^{\pm} = \frac{1}{2} \big(\tau_x \pm i\tau_y \big)$, and $\xi_{\k} = k^2/2m -\mu$ is the single-particle energy measured with respect to the Fermi level. We have also combined fermionic/bosonic Matsubara frequencies and momenta into the four-vectors $k = (i\w_n,\v{k})$ and $q = (i\W_m,\v{q})$, where $q$ corresponds to the center-of-mass momentum of the fermion pair and $k$ corresponds to the relative momentum.

We assume $g_1,g_2$ are such that we find a saddle-point configuration in which both $\Delta^{(1)}$ and $\Delta^{(2)}$ are condensed with a relative phase of $\pi/2$, breaking time reversal symmetry as discussed above. It will be convenient to change basis from $(\Delta^{(1)},\Delta^{(2)}) \goes (\Delta^+, \Delta^-)$ according to $\Delta^{(1)}_q \chi^{(1)}_{\k} + \Delta^{(2)}_q \chi^{(2)}_{\k} = \Delta^+_q \chi^+_{\k} + \Delta^-_q \chi^-_{\k}$ where the $\pm$ form factors are defined as
\begin{equation}
    \chi^\pm_{\k} = \eta_1 \, \chi^{(1)}_{\k} \pm i\,\eta_2 \,\chi^{(2)}_{\k}
\end{equation}
and $\eta_{1,2}$ quantify the relative magnitude of each order parameter component. We choose to normalize them such that $\eta_1^2 + \eta_2^2 = 1$ and can express both in terms of a ``mixing angle'' as $\eta_1  = \cos \eta$ and $\eta_2 = \sin \eta$.

Expanding around the saddle point with $\Delta^+_q = \Delta_0 = \text{const.}$ and $\Delta^-_q = 0$, the mean field equations are
\begin{align}
    \left( \frac{\eta_1^2}{g_1} + \frac{\eta_2^2}{g_2} \right) &= -T\sum_{k} \frac{|\chi_{\k}^+|^2}{(i\w_n)^2 - E_{\k}^2} \label{mfeq1} \\
    \left( \frac{\eta_1^2}{g_1} - \frac{\eta_2^2}{g_2} \right) &= -T\sum_k \frac{\left(\chi_{\k}^+ \right)^2}{(i\w_n)^2 - E_{\k}^2} \label{mfeq2}
\end{align}
where $E_{\k}^2 = \xi_{\k}^2 + |\Delta_{\k}|^2$ and the angle-dependent quasiparticle gap function is $\Delta_{\k} = \Delta_0 \chi_{\k}^+$. Given a particular set of pairing symmetries and coupling constants, these equations can be solved to determine the equilibrium values of $\eta$ and $\Delta_0$ which characterize the condensate. 

\textbf{\textsf{Effective action for fluctuations.}} 
Now, we move on to consider the fluctuations around this saddle point, which we parameterize as
\begin{equation}
\begin{split} \label{mode-defs}
    \Delta^+(x) &= \e^{2i\theta(x)} \big(\Delta_0 + h(x) \big) \\
    \Delta^-(x) &= \e^{2i\theta(x)} \big( a(x) + ib(x) \big)
    \end{split}
\end{equation}
where $\theta$ is the ABG phase mode, $h$ is the Higgs mode, and the $a$ and $b$ modes are fluctuations in the relative amplitude and phase of the two order parameter components, i.e. the generalized clapping modes. This parameterization suggests that we can equivalently think of the generalized clapping modes as being exciton-like fluctuations in the degenerate time-reversed $\Delta^-$ pairing channel, similar to Bardasis-Schreiffer modes \cite{bs-modes,jon-bs}, as illustrated in Fig. \ref{schematic-fig}(b).

To organize our calculations in a manifestly gauge-invariant way, we minimally couple the system to an external (classical) gauge field and perform a unitary transformation $\mathbb{G}^{-1} \goes U\mathbb{G}^{-1}U^\dag$ with $U = \e^{-i\theta \, \tau_z}$ such that the ABG mode and gauge field only appear together as the gauge-invariant vector field $V^0 = A^0 + \del_\tau \theta$ and $\v{V} = \v{A} - \v{\del} \theta $ (where we have set the electron charge equal to one). To quadratic order in these fields and setting $\v{q} \goes 0$, the action is
\begin{equation} \label{action-maintext}
    \begin{split}
        S &= \sum_q \left[\; \Pi^{00} V_{-q}V_q + n_s^{ij} \,V_{-q}^i V_{q}^j \right] \\
        &-\sum_q \left[ \Df^{-1}_h \, h_{-q}h_q + \Df^{-1}_a \, a_{-q}a_q + \Df^{-1}_b \, b_{-q}b_q \right] \\
        &+ \sum_q \left[\; \tilde{\Pi}^{ha} \, h_{-q}a_q + \Pi^{0b} V_{-q}^0 b_q \right]\,.
    \end{split}
\end{equation}
In the above, $\Pi^{00}$ is the electronic compressibility, $n_s^{ij}$ is the superfluid density,  $\Df_{h,a,b}$ are the propagators for the Higgs, relative amplitude, and relative phase modes, and $\tilde{\Pi}^{ha}$ and $\Pi^{0b}$ are linear couplings between the Higgs/relative amplitude and ABG/relative phase modes which are non-vanishing in the $\v{q} \goes 0$ limit. This action is derived and the correlation functions which appear in it are evaluated in appendix B. 

\begin{figure}
    \centering
    \includegraphics[width=80mm]{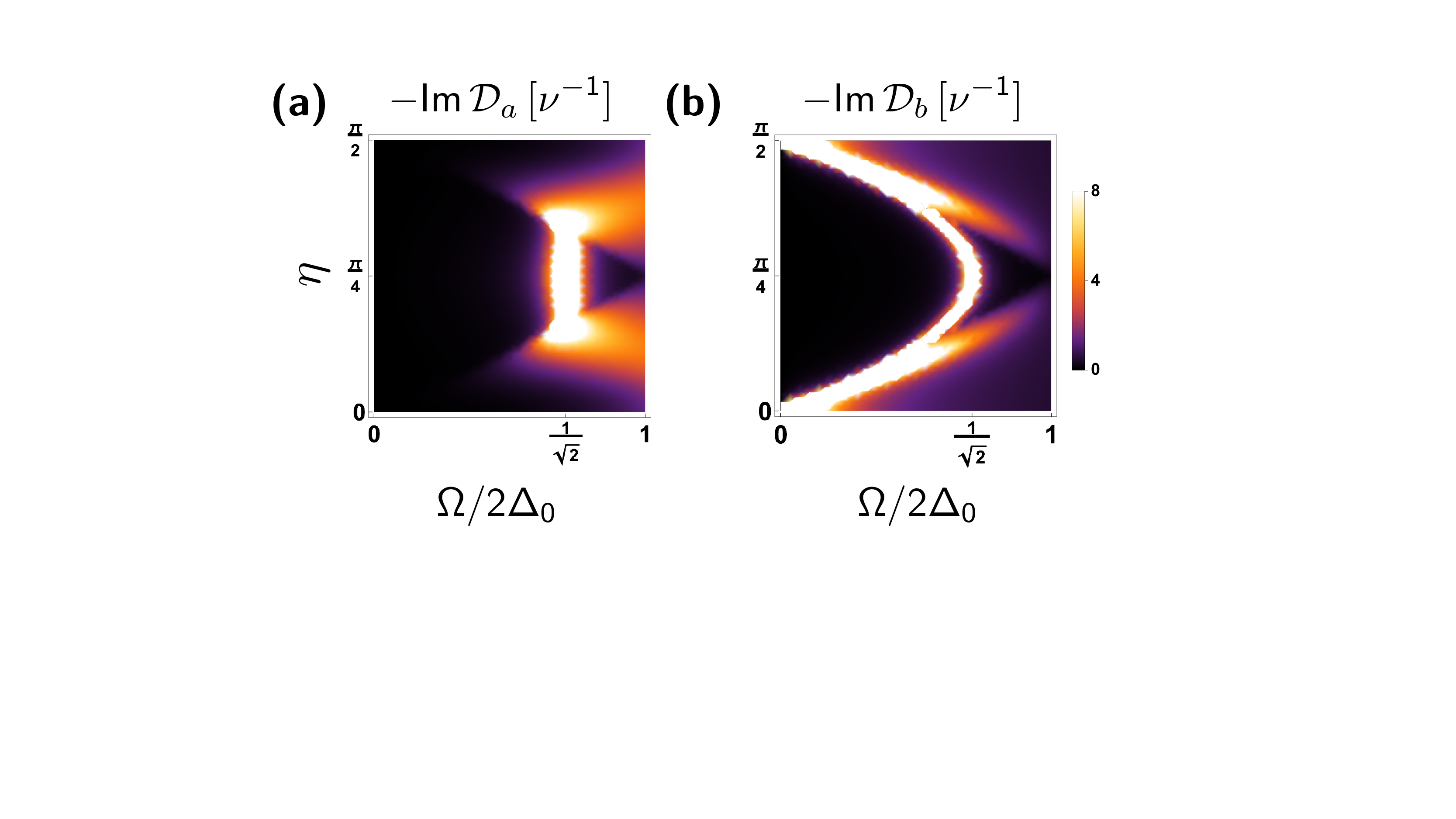}
    \caption{\textbf{$p$-wave superconductor. } Spectral functions (in units of the inverse normal state density of states at the Fermi level, $\nu^{-1}$) of \textbf{(a)} the relative amplitude and \textbf{(b)} phase modes for $p+ip$ pairing as a function of the mixing angle $\eta$. Note that for an equal admixture of the $p_x$ and $p_y$ components ($\eta = \pi/4$) the two modes are degenerate at the frequency $\sqrt{2}\Delta_0$. Away from this point, the modes split with $a$ moving toward the gap edge (taking the character of an amplitude mode), and $b$ moving toward zero frequency (taking the character of a phase mode). 
    }
    \label{p+ip-fig}
\end{figure}

\section*{\textsf{Results}}
To gain intuition, we begin by considering the familiar case of $p+ip$ pairing in two dimensions. When the $p_x$ and $p_y$ components of the order parameter occur with equal amplitudes ($\eta = \pi/4$), as is dictated by symmetry in most cases of physical interest, we find that after analytically continuing to real time the propagators for both the $a$ and $b$ modes have a pole at $\W = \sqrt{2}\Delta_0$, i.e. the two modes are degenerate. So, we see that the usual clapping modes previously studied in $p+ip$ superconductors and $^{\text{3}}$He-A are indeed a special case of the generalized clapping modes $a$ and $b$ studied in this work. Moreover, at this point the gap is isotropic and the couplings $\tilde{\Pi}^{ha}$ and $\Pi^{0b}$ vanish so that the (generalized) clapping modes decouple from both quasi-particle excitations and other collective modes and thus are infinitely long-lived at zero temperature. 

\begin{figure*}
    \centering
    \includegraphics[width=150mm]{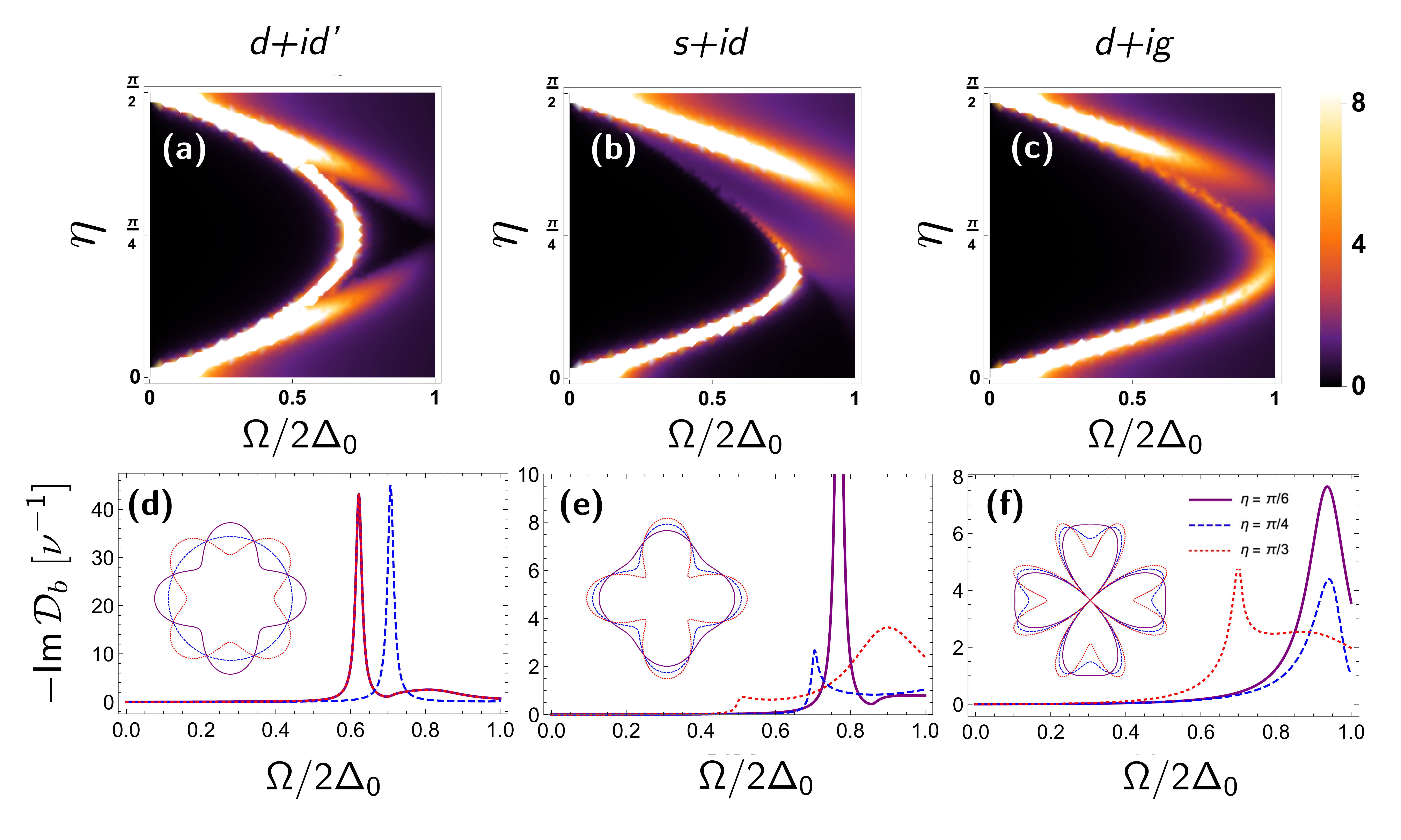}
    \caption{\textbf{Relative phase modes. } Spectral functions of the relative phase mode for all values of the mixing angle $\eta$ for \textbf{(a)} $d+id'$, \textbf{(b)} $s+id$, and \textbf{(c)} $d+ig$ orders. \textbf{(d,e,f)} Line cuts for several values of $\eta$ for each order parameter, with the corresponding quasiparticle gap function around the Fermi surface shown in the inset.}
    \label{2d-phase}
\end{figure*}

In real materials, anisotropic crystal fields can lead to deviations from the equal-amplitude $\eta=\pi/4$ state \cite{214-anisotropic}, making it interesting to consider the collective mode spectrum of the $p+ip$ state for general mixing angles. We plot the spectral functions of the $a$ and $b$ modes in Fig. \ref{p+ip-fig} as a function of the mixing angle $\eta$ and frequency $\W$. We see that the two modes split as the mixing angle deviates from $\pi/4$, so that for generic mixing angles there are two generalized clapping modes modes corresponding to the relative amplitude and phase fluctuations.

\textbf{\textsf{Two-dimensional systems. }} We will now investigate the generalized clapping mode spectrum for several even parity multi-component TRSB order parameters which are potentially relevant to experimental systems:  the $d_{x^2 - y^2} +id_{xy}$ state, originally studied in the context of the cuprate high-temperature superconductors \cite{laughlin-d+id,d+id-umd}, and now the subject of renewed interest due to its potential relevance to a number of moir\'e systems \cite{magic-angle-d+id,magic-angle-d+id-2,yakovenko-prx}, other heterostructures \cite{pi2jj}, and its proposed realization in twisted bilayers of cuprates \cite{twisted-bscco} and other unconventional superconductors \cite{volkov2020magic}; the $s+id_{xy}$ state, which has long been of interest in relation to the iron pnictide high-temperature superconductors \cite{fe-s+id, feas-s+id, hirschfeld-s+id}; and the $d_{x^2 -y^2} + ig_{xy(x^2 - y^2)}$ state which has recently been proposed as the order parameter of Sr$_{\text{2}}$RuO$_{\text{4}}$ \cite{kivelson-214,214-rus}. The basic properties of each order parameter studied in this work are listed in Table \ref{table}. 

Because these are all mixed symmetry states where the mixing angle is unconstrained by point group symmetries, it is important to survey the generalized clapping mode spectra over the full range of $\eta$. We first turn our attention to the relative phase mode, the spectra of which we plot for each of the above order parameters in Fig \ref{2d-phase}. Crucially, we observe that the relative phase mode resides below the quasiparticle continuum for all values of the mixing angle and, as shown in the line cuts in Fig. \ref{2d-phase}(d-f), remains coherent despite the fact that all of these order parameters possess point nodes for generic values of $\eta$.  

\begin{figure}
    \centering
    \includegraphics[width=80mm]{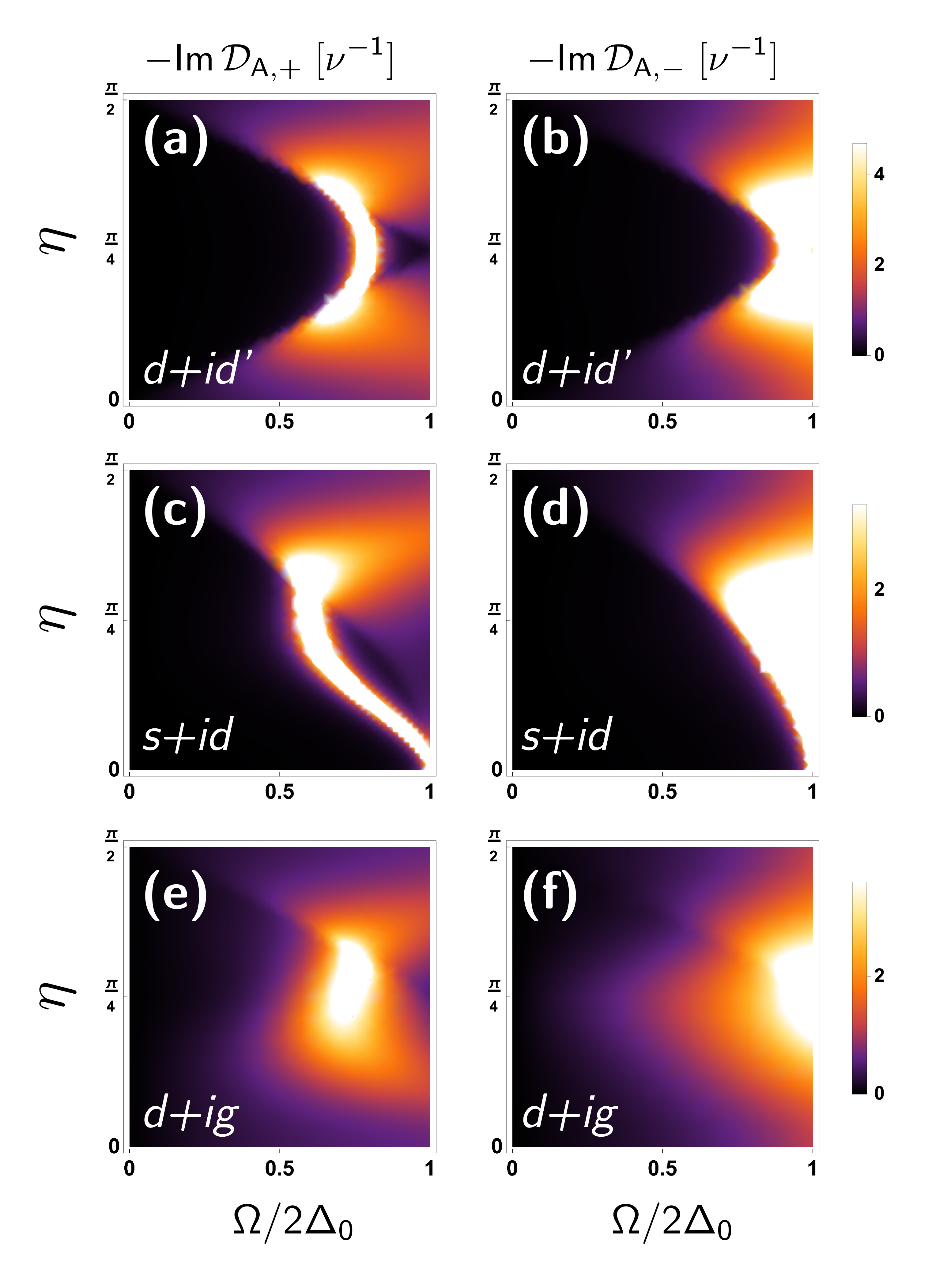}
    \caption{\textbf{Amplitude modes. } Spectral functions of the amplitude modes $A_{\pm}$ for \textbf{(a,b)} $d+ig$, \textbf{(c,d)} $d+id'$, and \textbf{(e,f)} $s+id$ orders.}
    \label{ampl-modes}
\end{figure}

Next, we turn our attention to the relative amplitude mode. As we show in appendix B, this mode is coupled to the Higgs mode even at zero momentum. The relative amplitude and Higgs modes then hybridize to form two orthogonal amplitude modes, which we call $A_+$ and $A_-$. We calculate the propagators $D_{A\pm}(\W)$ for these modes in appendix B, which we use to plot the spectral functions of each mode for various pairing symmetries, as shown in Fig. \ref{ampl-modes}. 

For all of the pairing symmetries studied, the $A_-$ mode resides at or slightly below the gap edge, much like the conventional Higgs mode in a single-component superconductor. More interesting is the $A_+$ mode, which lies well below the quasiparticle continuum for a wide range of mixing angles. This low-frequency amplitude mode represents a second novel collective excitation characteristic of TRSB superconducting states. 

We also note that the at $\eta = \pi/4$, the $d+id'$ state is fully gapped and chiral, and both generalized clapping modes are degenerate with a frequency of $\W = \sqrt{2}\Delta_0$, much like the chiral $p$-wave state in Fig. \ref{p+ip-fig} \cite{universalchiral}. As we will show below, this degeneracy is a general feature of chiral order parameters.

\textbf{\textsf{Three-dimensional systems}} We now consider several TRSB superconducting states in three-dimensional systems: the 3d $p_x + ip_y$ state (i.e. a $p_x +ip_y$ order parameter defined over a spherical Fermi surface), better known as the Anderson-Brinkman-Morel (ABM) state of $^{\text{3}}$He-A \cite{leggetrmp}, which is also a minimal model for a Weyl superconductor, and thus may bear some qualitative similarities to the superconducting state of UTe$_{\text{2}}$ \cite{ian-ute2} and other candidate Weyl systems; the three-dimensional $d_{x^2-y^2} +id_{xy}$ ``double Weyl'' superconducting state possibly realized in the ``hidden order'' phase of URu$_{\text{2}}$Si$_{\text{2}}$ \cite{weyl-uru2si2-2,weyl-uru2si2} or SrPtAs \cite{srptas-2,srptas}; and finally the $d_{xz} + id_{yz}$ state which is a candidate order parameter for Sr$_{\text{2}}$RuO$_{\text{4}}$ as well as the hidden order phase of URu$_{\text{2}}$Si$_{\text{2}}$ \cite{eg-uru2si2,eg-uru2si2-2}. 

\begin{figure}
    \centering
    \includegraphics[width=85mm]{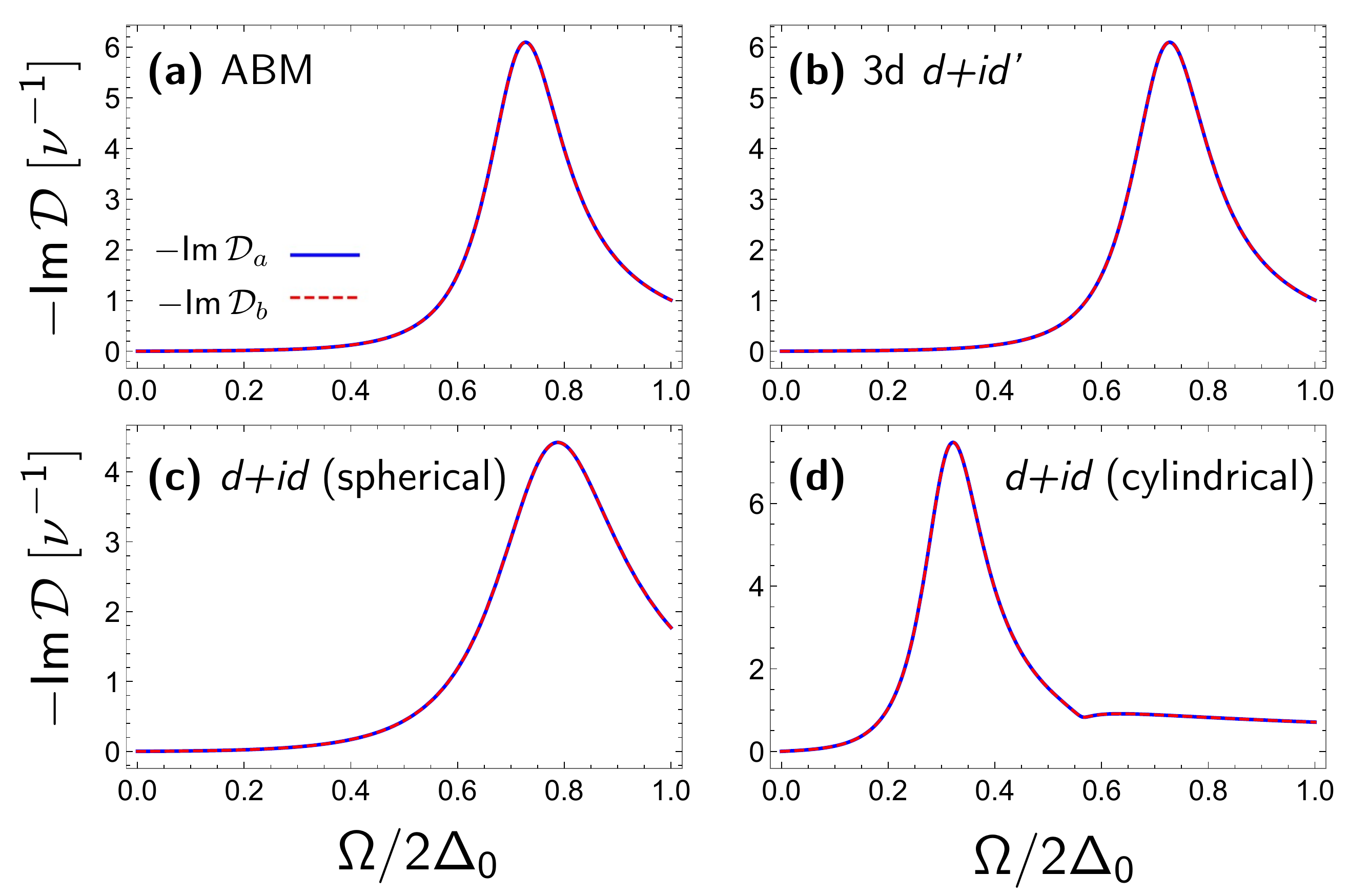}
    \caption{\textbf{Three-dimensional chiral superconductors. } Spectral functions for the relative amplitude and phase modes for the chiral ($\eta = \pi/4$) states \textbf{(a)} ABM, \textbf{(b)} three-dimensional $d+id'$, and $d+id$ state for a \textbf{(c)} fully three-dimensional system with a spherical Fermi surface, and \textbf{(d)} quasi-two-dimensional system with a cylindrical Fermi surface. In all cases, the two clapping modes are degenerate, reflecting the chiral nature of the order parameter.}
    \label{3d-fig}
\end{figure}

All of these states are most naturally considered as equal-admixtures, i.e. with $\eta = \pi/4$. As seen in Fig. \ref{3d-fig}, the generalized clapping modes are degenerate with one another in all cases. Unlike the two-dimensional case, however, the mode frequency is not $\sqrt{2}\Delta_0$. By inspection of the analytic form of the generalized clapping mode propagators (see appendix B), it is evident that the mode frequency is only $\sqrt{2}\Delta_0$ if the order parameter is both chiral {\it and} fully gapped, whereas the ABM and 3d $d+id'$ ($d+id$) states exhibit point (line) nodes. We also note that this demonstrates that the generalized clapping modes remain coherent even for a system with line nodes in the superconducting gap, i.e. these modes' survival is insensitive to the nodal structure of the order parameter. Combined with the prior results for the chiral $p+ip$ and $d+id'$ states in two-dimensions, we see that chiral order parameters are generically characterized by degenerate generalized clapping modes. In this way, the generalized clapping mode spectrum can be used to differentiate chiral states from mixed symmetry TRSB states, as further discussed below.

Altogether, the results we have presented so far provide strong evidence that the generalized clapping modes are always well-defined sub-gap excitations in TRSB superconducting states. A rigorous proof of this conjecture would be an interesting direction for future research, but is beyond the scope of the present work. At a practical level, this universality is essential to these modes' application in collective mode spectroscopy.

\section*{\textsf{Experimental detection schemes}}
Having established that the relative phase and amplitude modes are generically well-defined collective excitations in TRSB superconductors, we now propose several means for their experimental detection.

\begin{figure}
    \centering
    \includegraphics[width=90mm]{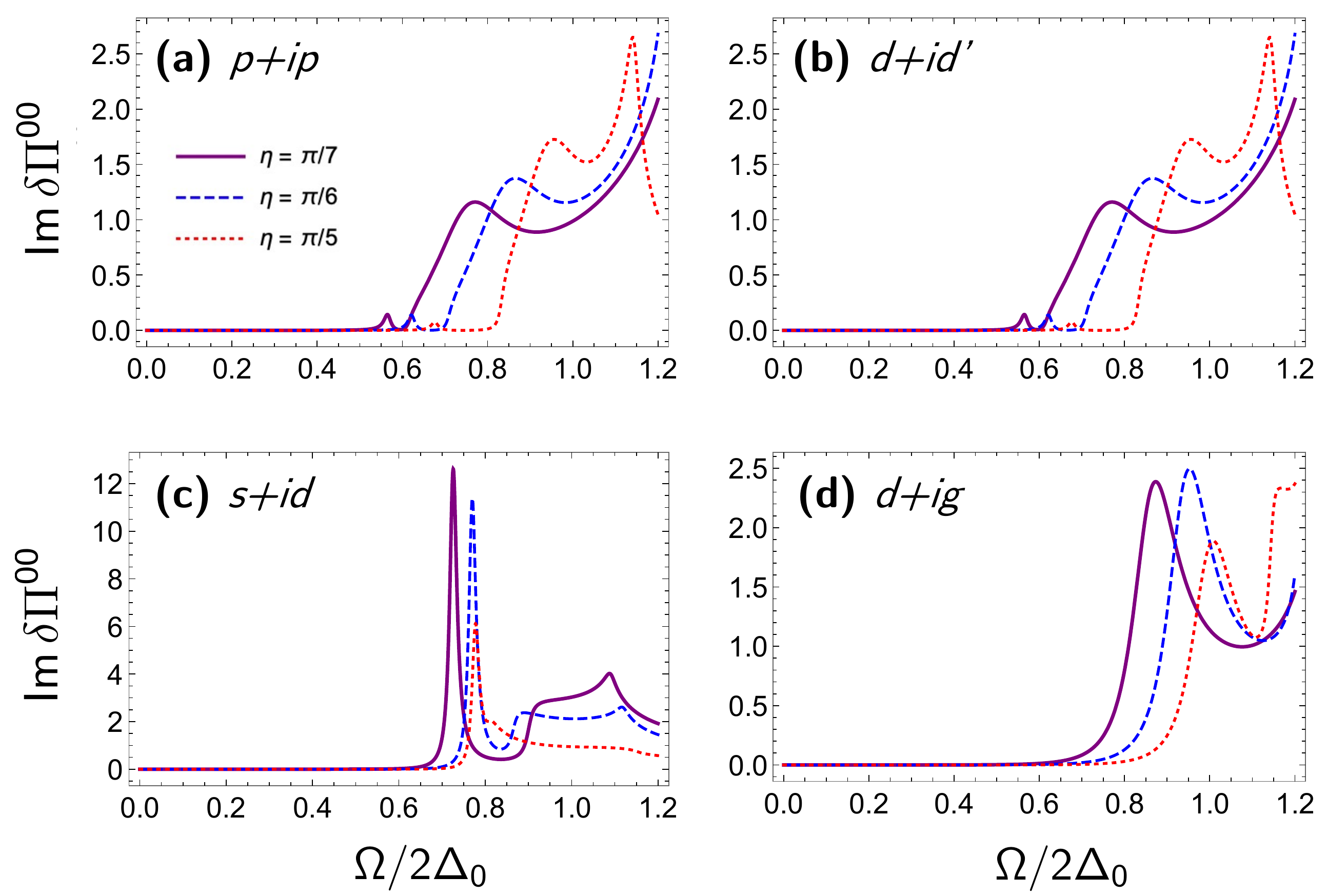}
    \caption{\textbf{ac compressibility. } Imaginary part of the relative phase mode's contribution to the electronic compressibility for \textbf{(a)} $p+ip$, \textbf{(b)} $d+id'$, \textbf{(c)} $s+id$, and \textbf{(d)} $d+ig$ orders, in units of $e^2\nu$. All exhibit clear features at the clapping mode frequency. }
    \label{compress-fig}
\end{figure}

\textbf{\textsf{Relative phase mode. }} From the action (\ref{action-maintext}), we see that the relative mode contributes to the charge density as $\rho \sim i\del_t b$ due to the non-vanishing linear coupling $\Pi^{0b}$ between $b$ and the scalar potential. One may then integrate $b$ out of this action, renormalizing the electronic compressibility $\Pi^{00}(\W) \goes \Pi^{00}(\W) + \delta \Pi^{00} (\W)$ where
\begin{equation}
     \delta \Pi^{00}(\W) =  - \Pi^{0b}(-\W) \, \Df_b(\W) \, \Pi^{0b}(\W)/4
\end{equation}
is the relative phase mode's contribution to $\Pi^{00}(\W)$. We plot this function for several pairing symmetries in Fig. \ref{compress-fig}, where we see clear features at the relative phase mode frequency. 

For most candidate TRSB superconductors, the superconducting gap, and hence the relative phase mode frequency, is in the terahertz regime. The ac electronic compressibility can be directly measured at THz frequencies using existing experimental techniques such as momentum-resolved electron energy loss spectroscopy (M-EELS) \cite{m-eels-review,matteo-compress,m-eels-2,m-eels-214}, which enables  direct experimental detection of the relative phase mode. 

In such an experiment, the relative phase mode can be distinguished from trivial non-electronic modes by its dispersion and the fact that the peak corresponding to this mode should vanish above $T_c$. It can be shown that the relative phase mode disperses as $\W^2 = \W_{\text{b}}^2 + \alpha v_F^2 q^2$, where $\W_{\text{b}}$ is the frequency of the mode and $\alpha$ is a constant dependent on the mixing angle and orbital symmetries of the order parameter. In light of this, the observation of a sub-gap peak in the charge response, measured via M-EELS, which disperses with an electronic-scale wavelength would constitute smoking gun evidence for the relative phase mode in a TRSB superconductor.

Alternatively, these modes may also be detected by sub-gap peaks in the  microwave power absorption, as shown in Refs. \onlinecite{power-1,power-2}. In fact, this technique has been used to observe a collective mode of unknown origin in the heavy-fermion superconductor UBe$_{\text{13}}$ \cite{ube13-cm}. To date, there is no evidence that the superconducting state of this system is TRSB, but a split transition has been reported in specific heat measurements of its Th-doped relative U$_{\text{1}-x}$Th$_x$Be$_{\text{13}}$ \cite{thube13-c,Stewart2019}, which is suggestive of a multi-component order. In light of this, it could be interesting to revisit the order parameter symmetry of these compounds and whether time-reversal symmetry is broken. If so, Ref. \onlinecite{ube13-cm} might represent the first measurement of a generalized clapping mode. 

In addition to the equilibrium probes discussed above, the relative phase mode should also be detectable out of equilibrium using time-resolved THz spectroscopy \cite{noneq-leggett}. In fact, a recent experiment has claimed to observe the Leggett phase mode in the multi-band superconductor MgB$_{\text{2}}$ using this technique \cite{thz-leggett-exp}. However, another THz study of MgB$_2$ attributed the same experimental signatures to other modes, complicating the identification of the Leggett mode \cite{mgb2-exp-arxiv}. In any case, further work is necessary to establish the precise experimental signatures of the relative phase mode in such an experiment.

\textbf{\textsf{Amplitude modes. }} The $A_{\pm}$ modes couple to electromagnetic fields in the same manner as the Higgs mode in a conventional single-component superconductor, through the nonlinear coupling $\delta S \sim \v{A}^2 \, A_{\pm}$. One method to detect amplitude modes using this coupling is via the observation of resonantly enhanced third harmonic generation in a nonlinear THz spectroscopy experiment, which has been successfully performed for both conventional \cite{nbn-thg} and high-temperature superconductors \cite{cuprate-thg}. Amplitude modes have also been detected in THz pump, optical probe experiments for conventional superconductors \cite{nbn-pump-probe} and cuprates \cite{bscco-pump-probe}. However, despite the rapid development of THz spectroscopy of a probe of Higgs modes in superconductors, a number of questions and controversies remain open. In particular, the impact of charge density fluctuations \cite{higgs-cdf-1}, disorder \cite{higgs-disorder-1,higgs-disorder-2,higgs-disorder-3}, and other modes \cite{thz-josephson-plasmons} on the THz response continue to be actively discussed, and make the identification of the Higgs mode in these experiments controversial. 

The $A_+$ mode in TRSB superconductors is attractive on account of its low frequency, which should make it less heavily damped than conventional Higgs modes and easier to disentangle from the charge density fluctuations which onset at the gap edge \cite{higgs-cdf}. Nonetheless, in light of the aforementioned controversy surrounding the THz detection of conventional Higgs modes, further work is necessary to establish the detailed THz response of the amplitude modes discussed in our work to facilitate comparison with potential future experiments.

Alternatively, the amplitude modes can be detected using microwave spectroscopy, or within linear response as a resonance in the optical conductivity in the presence of a background dc supercurrent \cite{higgs-supercurrent-theory}, as has been demonstrated in NbN \cite{nbn-supercurrent}. 

Like the relative phase mode discussed above, the observation of the low-frequency $A_+$ amplitude mode is a direct signature of the TRSB superconducting state. Thus, the detection of this mode represents a new avenue in the emerging field of ``Higgs spectroscopy'' \cite{cuprate-thg,non-eq-higgs-expt,higgs-spect-th} where non-equilibrium amplitude oscillations are used to gain insight into the symmetry of the condensate. 

\section*{\textsf{Discussion}}
To summarize, we have shown that fluctuations in the relative amplitude and phase of multi-component TRSB order parameters are generically well-defined collective modes with frequencies below the quasiparticle continuum. Moreover, even for nodal gap functions, we have found that these modes are not overdamped by low-energy quasiparticles in the $T\goes 0$ limit. The frequency of each mode depends strongly on both the orbital symmetries of the two order parameter components and the relative amplitude of the two components in equilibrium. 

Further, we have proposed a number of means to experimentally detect generalized clapping modes: the relative phase mode can be directly detected via measurement of the ac electronic compressibility (using e.g. M-EELS), the amplitude modes can be detected using ultrafast and non-linear THz spectroscopy as well as optical conductivity measurements (in the presence of a dc supercurrent), and both modes can be detected in microwave power absorption measurements. The observation of these modes in a given material would constitute robust, ``smoking gun'' evidence of a multi-component TRSB order parameter. 

\textbf{\textsf{Collective mode spectroscopy. }} Our work enables a variety of existing experimental techniques including ultrafast and non-linear optics, electron scattering, and microwave spectroscopy to be used as direct probes of TRSB in unconventional superconductors. These measurements can be thought of as a form of ``collective mode spectroscopy,'' where one obtains information about the structure of the order parameter from its collective mode spectrum. Given the relative scarcity of probes directly sensitive to TRSB in superconductors (previously limited to only Kerr rotation and muon spin relaxation), this represents a substantial expansion of the experimental tools available to characterize these exotic TRSB superconducting states. 

Detection of the generalized clapping modes also offers the unique ability to estimate the relative magnitudes of each order parameter component through the frequency at which the mode resides (as seen in all of our results, the phase mode frequency is maximum for an equal admixture of order parameter components, and decreases as one component or the other becomes dominant). Moreover, observing these modes using ultrafast THz techniques would be the first probe able to assess TRSB superconductivity in driven non-equilibrium superconductors.

\textbf{\textsf{Application to Sr$_{\textsf{2}}$RuO$_{\textsf{4}}$. }} As mentioned above, the generalized clapping mode spectrum can be used to unambiguously distinguish chiral TRSB states from non-chiral states, as a generic TRSB superconductor has two non-degenerate generalized clapping modes, whereas it is only in the special case of a chiral state that the two modes are degenerate. 

This unique capability is particularly well-suited to clarify the structure of the order parameter in Sr$_{\text{2}}$RuO$_{\text{4}}$ -- a problem of tremendous current interest. At the time of writing, the two leading candidate order parameters for this system are the mixed symmetry $d+ig$ state (see Fig. \ref{2d-phase}(c,f) and \ref{ampl-modes}(e,f)) and the chiral $d+id$ state (see Fig. \ref{3d-fig}(d), where the generalized clapping mode spectrum is plotted for the cylindrical Fermi surface relevant to Sr$_{\text{2}}$RuO$_{\text{4}}$). Quasiparticle interference \cite{214-qpi} and thermal transport \cite{214-th-cond} measurements have demonstrated that the order parameter exhibits vertical line nodes along the zone diagonals (i.e. along the $[110]$ direction), consistent with the $d+ig$ scenario, and counter to the horizontal line nodes expected for a $d+id$ state. In contrast, recent muon spin relaxation measurements failed to detect a splitting of the critical temperature under hydrostatic pressure \cite{214-hydro-musr}, which is only consistent with the chiral $d+id$ order.
Given these seemingly conficting results, a measurement of the generalized clapping mode spectra, via e.g. microwave absorption measurements, could provide crucial insight into the chirality, or lack thereof, of the order parameter, and help the community converge on one candidate order parameter over the other.

\textbf{\textsf{Outlook. }} Beyond its utility as a novel form of spectroscopy, the detection of generalized clapping modes is also interesting from a fundamental physics perspective, as the analogue of the clapping mode in $^3$He-A has yet to be realized in any electronic system. Previously, the existence and detection of clapping modes was considered only in $p+ip$ superconductors \cite{214-clapping, p+ip-em, 214-1d}, which have proven elusive to realize experimentally \cite{214-nmr}. This work opens up a number of other materials platforms \cite{214-kerr,214-musr,upt3-kerr,upt3-musr,uru2si2-kerr,ute2-original,ian-ute2,prossb-kerr,pr-musr, pnictide,bini-kerr,jimmy,twisted-bscco,volkov2020magic,magic-angle-d+id,magic-angle-d+id-2} as candidate systems to finally realize these exotic collective modes in a solid state system.

Finally, we speculate that our work may also be extended to the enigmatic pseudogap phase of the cuprate high-temperature superconductors where a non-zero Kerr rotation has been been reported \cite{ybco-kerr}, suggesting the existence of a TRSB phase above the superconducting transition. It has recently been suggested \cite{trsbmetal-1,trsbmetal-2,yakovenko-prx,2021phasefluctuation,k-122-metalexp} that the relative phase between two order parameter components can acquire a phase stiffness before either order parameter becomes phase coherent and condenses. In this scenario, the relative phase mode studied in this work could persist even above the superconducting transition, representing a truly novel collective excitation in a TRSB metallic phase.

\begin{acknowledgments}
The authors thank Matteo Mitrano and Eugene Demler (Harvard), Dmitri Basov (Columbia), Manfred Sigrist (ETH Z\"urich), and Roman Lutchyn (Station Q) for insightful discussions about this work. We also thank Charlotte B\o ttcher, Marie Wesson, Uri Vool, Yuval Ronen, Zachary Raines, Andrew Allocca, and Zhiyuan Sun for fruitful discussions through various iterations of this study. This work is primarily supported by the Quantum Science Center (QSC), a National Quantum Information Science Research Center of  the  U.S.  Department  of  Energy  (DOE). N.R.P. is supported by the Army Research Office through an NDSEG fellowship. J.C. is an HQI Prize Postdoctoral Fellow and gratefully acknowledges support from the Harvard Quantum Initiative. A.Y. is partly supported by 
the Gordon and Betty Moore Foundation through Grant GBMF 9468 and by the National Science Foundation under Grant No. DMR-1708688. P.N. is a Moore Inventor Fellow and gratefully acknowledges support through Grant GBMF8048 from the Gordon and Betty Moore Foundation.
\end{acknowledgments}

\appendix 

\section{TRSB superconductivity} \label{trsb-sec}
Throughout this work, we will consider a superconducting state characterized by a multi-component order parameter $\Delta = (\Delta_1, \Delta_2)$, where the two order parameter components $\Delta_1$ and $\Delta_2$ generically have different orbital symmetries. The most common scenario in which such a multi-component state is realized is in the case of a multi-band superconductor, where electrons on one band couple to $\Delta_1$, while the electrons on a second band couple to $\Delta_2$. However, it is also possible to have a multi-component state where all of the electrons in the system couple to both $\Delta_1$ and $\Delta_2$. The  most natural way this occurs is if the leading superconducting instability is in a symmetry channel belonging to a multi-dimensional irreducible representation of the crystalline point group. For example, the $E_u$ representation of the tetragonal point group has two basis functions $\{ k_x, k_y\}$ which must condense at the same temperature by symmetry. The order parameter will then have the multi-component form $\Delta = \Delta_1 + \Delta_2$, where $\Delta_1$ has the symmetry of the $k_x$ basis function and $\Delta_2$ has the symmetry of $k_y$. 

Alternatively, this kind of multi-component state can also come about due to an accidental degeneracy between pairing in two channels belonging to different irreducible representations of the point group. That is, one can have a ``mixed symmetry'' state with the multi-component order parameter $\Delta = \Delta_1 + \Delta_2$, where $\Delta_1$ and $\Delta_2$ belong to two different irreducible representations. This situation, although finely tuned, is potentially relevant to understanding the superconducting state of Sr$_{2}$RuO$_4$ \cite{kivelson-214,214-rus}. 

Finally, a mixed symmetry multi-component state can also be realized by way of two successive superconducting transitions. That is, rather than pairing in two channels being exactly degenerate, if there are two channels which are \emph{nearly} degenerate, the system can first condense in the dominant channel, and subsequently undergo another transition where the second component condenses as the temperature is further lowered. Below the lower transition, a multi-component state is realized. This scenario is likely applicable to UPt$_3$ \cite{upt3-kerr} and UTe$_2$ \cite{ian-ute2} where two superconducting transitions have been observed. 

We can begin by studying a multi-component state at a phenomenological level by constructing the general form of the Ginzburg-Landau free energy functional. Assuming a time-reversal invariant normal state and neglecting spin-orbit coupling, this functional reads
\begin{equation}
    \begin{split} \label{gl}
        f_{\text{GL}} &= \alpha_1 |\Delta_1|^2 + \alpha_2 |\Delta_2|^2 + \gamma \big(\bar{\Delta}_1 \Delta_2 + \Delta_1 \bar{\Delta}_2 \big)  \\[4pt]
        &+ \beta_{11}|\Delta_1|^4 + \beta_{22}|\Delta_2|^4 + \beta_{12}|\Delta_1|^2 |\Delta_2|^2 \\[4pt]
        &+ \lambda \big(\bar{\Delta}_1^2 \Delta_2^2 + \Delta_1^2 \bar{\Delta}_2^2 \big)\, .
    \end{split}
\end{equation}
Minimizing $f_{\text{GL}}$ with respect to $\Delta_1$ and $\Delta_2$ will determine the equilibrium configuration of the condensate. For a multi-component order parameter, this corresponds to determining the magnitudes $|\Delta_1|$ and $|\Delta_2|$ as well as the relative phase $\varphi$ between the two components (recall that the free energy is independent of the overall phase of the condensate). 

Let us first consider the case of multi-band superconductivity, where the two order parameter components belong to different electronic bands. The last term in the first line of Eq. (\ref{gl}) then represents an interband Josephson coupling between the two order parameter components, which will dominate over the second-order Josephson coupling in the last line. The relative phase is then chosen to minimize this contribution to the free energy, so that in equilibrium $\varphi = 0$ if $\gamma < 0$ and $\varphi = \pi$ if $\gamma >0$. In either case, the equilibrium configuration of the order parameter is time-reversal invariant. 

Given the relative phase is pinned to a fixed value, one expects the collective mode spectrum to include a massive mode corresponding to fluctuations of $\varphi$ around $0,\pi$. This expectation is correct, and the mode in question is the well-known Leggett mode \cite{leggett-mode}, the properties of which have been thoroughly established from microscopic calculations \cite{leggett-review}. 

Next, we consider the remaining scenarios which lead to a multi-component order parameter. If $\Delta_1$,$\Delta_2$ belong to a multi-dimensional irreducible representation, we must have $\alpha_1 = \alpha_2$ and $\beta_{11} = \beta_{22}$, which immediately implies that $|\Delta_1| = |\Delta_2|$. For an accidental degeneracy between two irreducible representations, one has $\alpha_1 = \alpha_2$ but $\beta_{11} \neq \beta_{22}$ so that the magnitudes of each order parameter component need not be equal. The same is true for an accidental near-degeneracy where $\alpha_1 \approx \alpha_2$. 

\begin{figure}
    \centering
    \includegraphics[width=80mm]{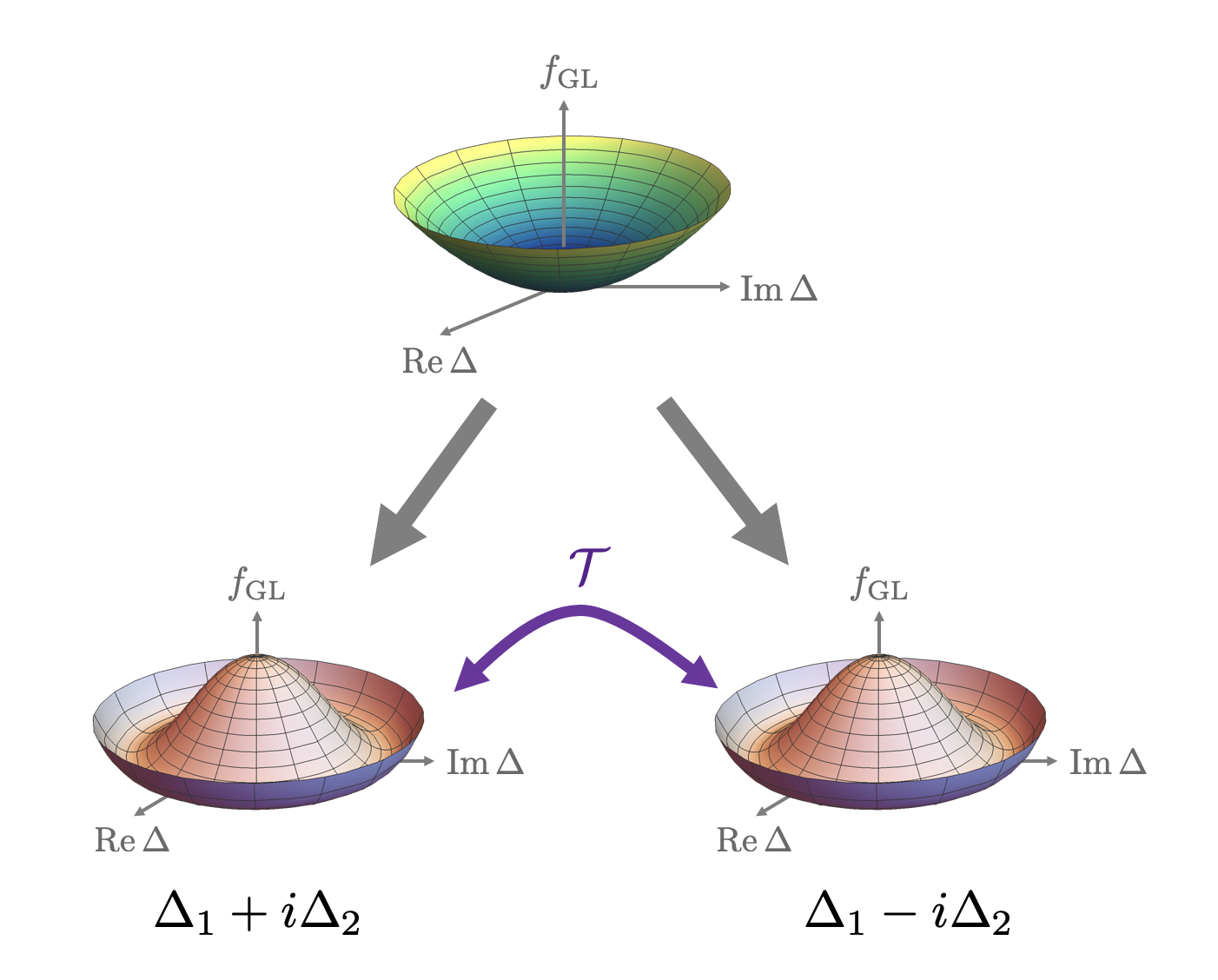}
    \caption{Illustration of TRSB superconducting ground states. At the superconducting transition, two degenerate ground states $\Delta_1 \pm i\Delta_2$ emerge, related by the time-reversal operator $\mathcal{T}: i \goes -i$. When the system condenses into one ground state or the other, time-reversal symmetry is spontaneously broken in addition to global $U(1)$ symmetry.}
    \label{trsb-gs}
\end{figure}

In all three cases, $\Delta_1$ and $\Delta_2$ have different symmetries, and thus there are no symmetry-allowed linear couplings between them, i.e. $\gamma = 0$. The leading order phase-sensitive term is then the second-order Josephson coupling in the last line of Eq. (\ref{gl}), which is minimized for $\varphi = \pm \pi/2$ if $\lambda > 0$, and $\varphi = 0,\pi$ if $\lambda < 0$.

In both cases, there are two degenerate minima, indicating that the order parameter breaks a discrete symmetry in addition to global $U(1)$. In the former case, the order parameter takes the form $\Delta = \Delta_1 \pm i\Delta_2$, corresponding to the TRSB superconducting state which is the subject of our interest in this work, and is illustrated schematically in Fig. \ref{trsb-gs}. As mentioned in the main text, a number of real materials are believed to realize this exotic state. 

Before moving on, we note that the other possibility, with $\lambda < 0$, gives rise to a nematic superconducting state where the magnitude of the order parameter $\Delta = \Delta_1 \pm \Delta_2$ breaks discrete crystalline symmetries. This state is much less common than the TRSB state, with the only current candidate systems being Cu-doped Bi$_2$Se$_3$ \cite{fu-nematic,cubise-nmr,nematic-stm} (although at higher dopings this system a TRSB superconducting state might also develop \cite{cubise-chiral}) and possibly magic angle twisted bilayer graphene \cite{tmablg}. We also note in passing that a different kind of TRSB superconducting state can be realized in a multi-band system in the interesting special case where there are three or more bands with frustrated interband couplings \cite{valentin-1,valentin-2,three-band-4,multiband-dia,pnictide-1,pnictide-2,pnictide-4}.

\begin{figure}
    \centering
    \includegraphics[width=85mm]{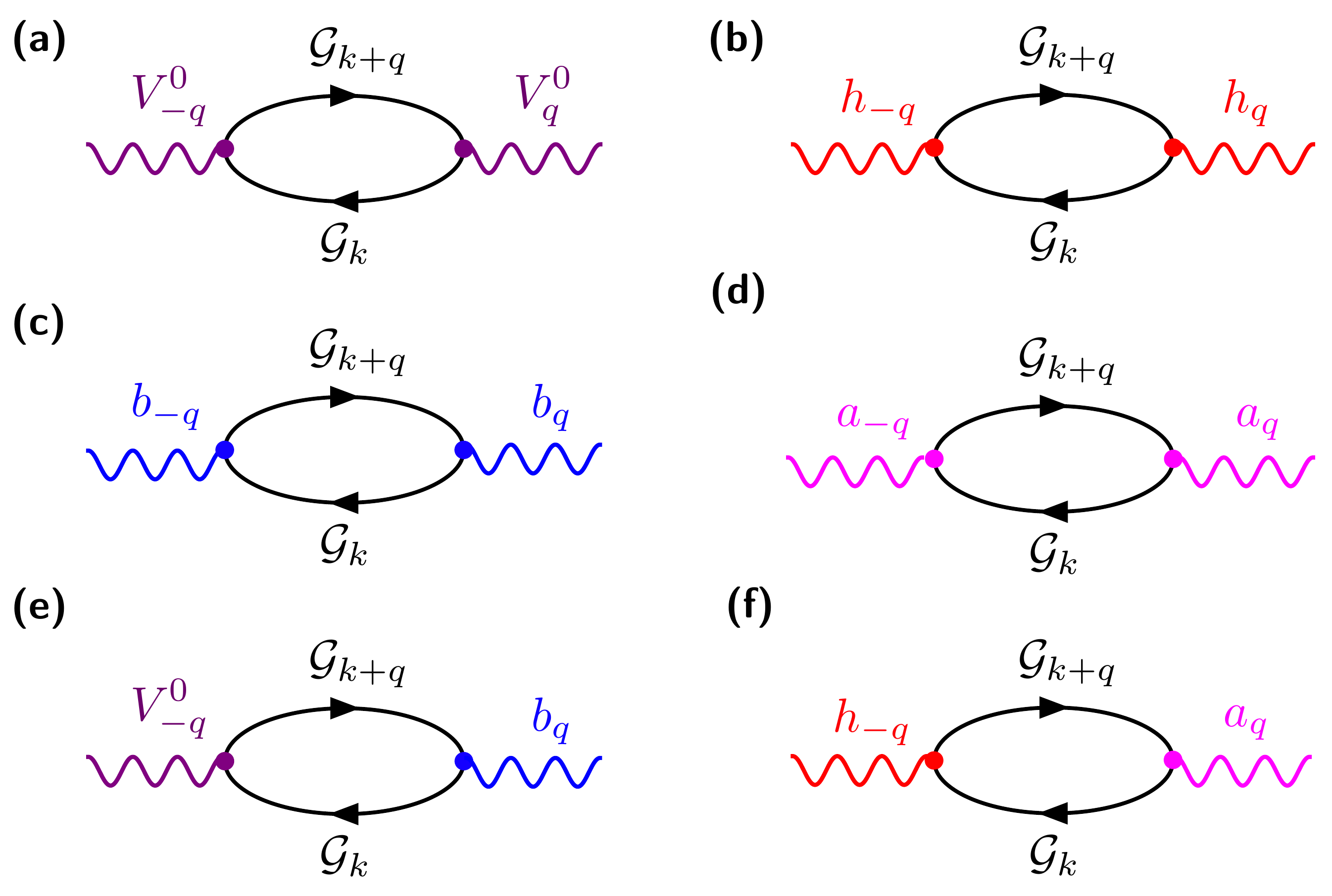}
    \caption{Bubble diagrams which contribute to the (a) electronic compressibility, $\Pi^{00}$; (b) Higgs propagator, $\Df_h^{-1}$; (c) relative phase mode propagator, $\Df_b^{-1};$ (d) relative amplitude mode propagator, $\Df_a^{-1}$; (e) linear coupling between the scalar potential and relative phase mode, $\Pi^{0b}$; (f) linear coupling between the Higgs and relative amplitude mode, $\tilde{\Pi}^{ha}$.}
    \label{bubbles-fig}
\end{figure}

\section{Calculation of the quadratic action} \label{bigappendix}
Here, we present the derivation of the effective action for the collective modes in a TRSB superconductor and evaluate the correlation functions that appear in it. Expanding the generic action for the order parameter (Eq. 1 in the main text) to quadratic order in the fluctuating fields defined in Eq. 5 of the main text about the $\Delta^+_q = \Delta_0; \;\Delta^-_q = 0$ saddle point, one finds
\begin{equation}
    \begin{split}
        S = &\left( \frac{\eta_1^2}{g_1} + \frac{\eta_2^2}{g_2} \right) \sum_q  \Big(2\Delta_0 h_0 + h_{-q}h_q + a_{-q}a_q + b_{-q}b_q  \Big) \\ 
        +&\left( \frac{\eta_1^2}{g_1} - \frac{\eta_2^2}{g_2} \right)\sum_q \Big( 2\Delta_0 a_0 + 2h_{-q}a_{q} \Big) - \tr \log \mathbb{G}^{-1} \, .
    \end{split}
\end{equation}
The full inverse fermion propagator is 
\begin{equation}
    \mathbb{G}^{-1}_{kk'} = \mathcal{G}^{-1}_k \, \delta_{kk'} - \mathbb{V}_{kk'}^{(1)} - \mathbb{V}_{kk'}^{(2)} - \mathbb{H}_{kk'} - \mathbb{C}_{kk'}
\end{equation}
where $\mathcal{G}_k^{-1} = i\w_n - \xi_{\k}\,\tau_z + \Delta_{\k}^+ \,\tau^+ + \bar{\Delta}_{\k}\,\tau^-$ and the vertices appearing above are
\begin{align}
     \mathbb{V}^{(1)}_{kk'} &= iV^0_{k-k'} \, \tau_z - \frac{1}{m} \, \v{V}_{k-k'}\cdot \pfrac{\v{k} +\v{k}'}{2} \\
    \mathbb{V}^{(2)}_{kk'} &= \frac{1}{2m} \, \v{V}_{k'-k} \cdot \v{V}_{k-k'} \, \tau_z \\
    \mathbb{H}_{kk'} &= -h_{k-k'} \, \chi^+_{(\k+\k')/2} \, \tau^+ - \bar{h}_{k'-k} \, \chi^-_{(\k+\k')/2} \, \tau^- \\
    \mathbb{C}_{kk'} &= -\Delta^-_{k-k'} \, \chi_{(\k+\k')/2}^- \, \tau^+ - \bar{\Delta}^-_{k'-k} \, \chi^+_{(\k+\k')/2} \, \tau^-
\end{align}
Next, we expand the functional logarithm to quadratic order in the fluctuating fields $V, h, a$, and $b$, leading to the effective action
\begin{widetext}
\begin{subequations}
\begin{align}
 S = &\left( \frac{\eta_1^2}{g_1} + \frac{\eta_2^2}{g_2} \right) \sum_q  \Big(2\Delta_0 h_0 + h_{-q}h_q + a_{-q}a_q + b_{-q}b_q  \Big) \label{hsactpt1}\\[4pt]
        +&\left( \frac{\eta_1^2}{g_1} - \frac{\eta_2^2}{g_2} \right)\sum_q \Big( 2\Delta_0 a_0 + 2h_{-q}a_{q} \Big)\label{hsactpt2} \\[4pt]
        +&\, \tr \left( \mathcal{G}\mathbb{V}^{(1)}\right) +  \tr \left( \mathcal{G}\mathbb{V}^{(2)}\right) + \frac{1}{2} \, \tr \left( \mathcal{G} \mathbb{V}^{(1)} \mathcal{G} \mathbb{V}^{(1)}\right) \label{abgact}\\[4pt]
        +& \, \tr \left(\mathcal{G} \mathbb{H} \right) + \frac{1}{2} \left(\mathcal{G}\mathbb{H}\mathcal{G}\mathbb{H}\right) \label{higgsact} \\[4pt]
        +& \, \tr \left( \mathcal{G} \mathbb{C}\right) + \frac{1}{2} \tr \left( \mathcal{G} \mathbb{C} \mathcal{G} \mathbb{C}\right)\label{clappingmodesact} \\[4pt]
        +& \, \tr \left( \mathcal{G}\mathbb{C} \mathcal{G} \mathbb{V}^{(1)} \right) + \tr \left( \mathcal{G} \mathbb{C} \mathcal{G} \mathbb{H} \right) + \tr \left( \mathcal{G} \mathbb{H} \mathcal{G} \mathbb{V}^{(1)}\right)  \, . \label{couplingsact}
\end{align}
\end{subequations}
Our task is now to evaluate each of these contributions, which when taken together determine the collective mode spectrum of the system. Throughout, we will focus on the long wavelength $\v{q} \goes 0$ limit which is most relevant to the experimental detection of, and coupling to, these collective modes.

\subsection{Electronic compressibility}
First, we will evaluate the terms appearing in (\ref{abgact}) which will give the dispersion of the ABG mode/gauge field. The first term is a total derivative that does not affect dynamics, and the second gives the diamagnetic contribution to the superfluid density. The last term can be broken into three contributions,
\begin{equation}
    \begin{split}
        \frac{1}{2} \tr \left( \mathcal{G} \mathbb{V}^{(1)} \mathcal{G} \mathbb{V}^{(1)} \right) &= -\frac{T}{2}\sum_{kq}\tr \left[\mathcal{G}_{k+q}\tau_z \mathcal{G}_k \tau_z\right]V^0_{-q}V^0_q \\ &+ \frac{T}{2m^2}\sum_{kq} \tr \left[ \mathcal{G}_{k+q}\mathcal{G}_k\right]k^i k^j \, V_{-q}^i V_q^j \\
        & \; \; \; - \frac{iT}{m}\sum_{kq} \tr \left[ \mathcal{G}_{k+q} \tau_z \mathcal{G}_k \right] \, k^i \, V^0_{-q}V^i_q  \\
        &\equiv \sum_q \Big(\Pi^{00}_q \, V^0_{-q}V^0_q + \Pi^{ij}_q \, V^i_{-q}V^j_q  + \Pi^{0i}_q \, V^0_{-q}V^i_q \Big) \, .
    \end{split}
\end{equation}
Here, $\Pi^{00}_q$ is the compressibility and $\Pi^{0i}_q$ is an ``anomolous'' contribution which is only present in TRSB systems. However, for an even-parity superconducting state, $\Pi^{0i}_q = 0$ for $\v{q} = 0$ by virtue of inversion symmetry, so we will neglect this term going forward. Finally, the correlator $\Pi^{ij}_q$ is related to the superfluid density by $n_s^{ij} = n\delta^{ij} + \Pi^{ij}_{q=0}$.

We begin by calculating the compressibility $\Pi^{00}(i\W_m)$ for $\v{q}=0$. Performing the Nambu trace, we have
\begin{equation}
    \Pi^{00}(i\W_m) = -\frac{T}{2}\sum_k \tr \left[\mathcal{G}_{k+q} \tau_z \mathcal{G}_k \tau_z \right] = -T\sum \frac{(i\w_n)^2 + i\w_n \,i\W_m + E_{\k}^2 -2|\Delta_{\k}|^2}{[(i\w_n + i\W_m)^2 - E_{\k}^2][(i\w_n)^2 - E_{\k}^2]} \, .
\end{equation}
The Matusbara sums can be performed, yielding
\begin{equation} \label{compressibility}
    \Pi^{00}(i\W_m) = -\int_{\k} \frac{\tanh(E_{\k}/2T)}{2E_{\k}} \; \frac{(2\Delta_0)^2 |\chi^+_{\k}|^2}{(i\W_m)^2 - (2E_{\k})^2} \, .
\end{equation}
Note that in the static limit for a single-component $s$-wave superconductor at zero temperature, this reduces to the standard result $\Pi^{00} = \nu$, where $\nu$ is the density of states at the Fermi level. 

\subsection{The Higgs mode}
Next, we move onto the terms (\ref{higgsact}) which give the dispersion of the Higgs mode. The linear term is 
\begin{equation}
    \tr \left(\mathcal{G}\mathbb{H} \right) = -h_0 T\sum_k \Big( \chi^+_{\k} \, \tr \left[\mathcal{G}_k \tau^+ \right] + \chi^-_{\k} \, \tr \left[ \mathcal{G}_k \tau^-\right] \Big) = 2\Delta_0 h_0 \, T\sum_k \frac{|\chi_{\k}^+|^2}{(i\w_n)^2 - E_{\k}^2}  \, .
\end{equation}
From the saddle point equation (Eq. 3 in the main text), this term cancels against the Hubbard-Stratonovich term in the action.

Next, we calculate the quadratic term, given by
\begin{equation}
    \frac{1}{2} \tr \Big(\mathcal{G} \mathbb{H}\mathcal{G} \mathbb{H} \Big) = \frac{T}{2}\sum_{kq} |\chi^+_{\k}|^2 \; \frac{(i\w_n)^2 - E_{\k}^2 + 2|\Delta_{\k}|^2 + i\w_n \, i\W_m - \xi_{\k} \,i\W_m}{[(i\w_n + i\W_m)^2 - E_{\k}^2][(i\w_n)^2 - E_{\k}^2]}\; h_{-q}h_q \equiv \sum_q \Pi^{hh}(i\W_m) \, h_{-q}h_q \, .
\end{equation}
Note that the last term which is linear in $\xi_{\k}$ vanishes by particle-hole symmetry. The surviving sums can be computed using the mean field equation (Eq. 3 in the main text), yielding 
\begin{equation}
   \Pi^{hh}(i\W_m) = -\left( \frac{\eta_1^2}{g_1} + \frac{\eta_2^2}{g_2} \right) + \int_{\k} \; |\chi^+_{\k}|^2 \; \frac{\tanh(E_{\k}/2T)}{2E_{\k}} \; \frac{(i\W_m)^2 - 4|\Delta_{\k}|^2}{(i\W_m)^2 - (2E_{\k})^2} \, .
\end{equation}
Combining with the bare term in (\ref{hsactpt1}), we find the quadratic action for the Higgs mode is $-\sum_q \Df_h^{-1}(i\W_m) \, h_{-q}h_q$ with the inverse propagator
\begin{equation}
    \Df_h^{-1}(i\W_m) = -\left( \frac{\eta_1^2}{g_1} + \frac{\eta_2^2}{g_2} \right)- \Pi^{hh}(i\W_m) =   -\int_{\k} \; |\chi^+_{\k}|^2 \; \frac{\tanh(E_{\k}/2T)}{2E_{\k}} \; \frac{(i\W_m)^2 - 4|\Delta_{\k}|^2}{(i\W_m)^2 - (2E_{\k})^2} \, .
\end{equation}
Note that in the limit of an $s$-wave superconductor at zero temperature, this mode has a gap of $2\Delta_0$, as expected.

\subsection{The generalized clapping modes}
Next, we consider (\ref{clappingmodesact}) to determine the dispersion of the relative amplitude and phase modes. First, we note that Eq. 4 in the main text has one particularly useful implication: since the left-hand side is real, the right-hand side must be as well, implying that $(\chi^+_{\k})^2 = (\chi^-_{\k})^2$ inside the angular integral. In fact, this holds for $(\chi^+_{\k})^2$ averaged against any real function.

With this fact in mind, the linear term in (\ref{clappingmodesact}) is 
\begin{equation}
\tr \left(\mathcal{G}\mathbb{C}\right) = -\Delta^-_0 T\sum_k \chi^-_{\k} \tr \left[ \mathcal{G}_k \tau^+\right] - \bar{\Delta}^-_0 T\sum_k\chi_{\k}^+ \, \tr \left[ \mathcal{G}_k \, \tau^- \right] = \Delta_0 T \sum_k \frac{\left( \chi^-_{\k} \right)^2 \, \Delta_0^- + \left(\chi_{\k}^+ \right)^2 \bar{\Delta}_0^- }{(i\w_n)^2- E_{\k}^2} \, .
\end{equation}
Using the mean field equation (Eq. 4 in the main text) and the decomposition $\Delta^-_q = a_q + ib_q$, this becomes
\begin{equation}
    \tr \left(\mathcal{G}\mathbb{C}\right) = \Delta_0 T\sum \frac{\big( \chi_{\k}^+\big)^2 }{(i\w_n)^2 - E_{\k}^2} \, \Big(\Delta^-_0 + \bar{\Delta}^-_0 \Big) = -2\left( \frac{\eta_1^2}{g_1} - \frac{\eta^2_2}{g_2} \right) \Delta_0 a_0 
\end{equation}
which cancels against the Hubbard-Stratonovich term in the action. Moving onto the quadratic term, we have
\begin{equation}
   \frac{1}{2}\,\tr \Big( \mathcal{G}\mathbb{C}\mathcal{G}\mathbb{C} \Big) = \sum_q\Big(\Pi^{aa}_q \, a_{-q}a_q + \Pi^{bb}_q \, b_{-q}b_q + 2 \Pi^{ab}_q \, a_q b_{-q} \Big) \, .
\end{equation}
Focusing on the $\v{q} = 0$ limit, the correlation functions above are 
\begin{align}
    \Pi^{aa}_q &= T\sum_k \, | \chi_{\k}^-|^2 \; \frac{[(i\w_n)^2 - E_{\k}^2] + i\w_n \, i\W_m + |\Delta_{\k}|^2}{[(i\w_n + i\W_m)^2 - E_{\k}^2][(i\w_n)^2 - E_{\k}^2]} + \frac{T\Delta_0^2}{2}\sum_k \frac{\left(\chi_{\k}^+ \right)^4 + \left(\chi_{\k}^- \right)^4}{[(i\w_n + i\W_m)^2 - E_{\k}^2][(i\w_n)^2 - E_{\k}^2]} \\
     \Pi^{bb}_q &= T\sum_k \, | \chi_{\k}^-|^2 \; \frac{[(i\w_n)^2 - E_{\k}^2] + i\w_n \, i\W_m + |\Delta_{\k}|^2}{[(i\w_n + i\W_m)^2 - E_{\k}^2][(i\w_n)^2 - E_{\k}^2]} - \frac{T\Delta_0^2}{2}\sum_k \frac{\left(\chi_{\k}^+ \right)^4 + \left(\chi_{\k}^- \right)^4}{[(i\w_n + i\W_m)^2 - E_{\k}^2][(i\w_n)^2 - E_{\k}^2]} \\
    \Pi_q^{ab} &= \frac{iT\Delta_0^2}{2} \sum_k \frac{\left(\chi_{\k}^+ \right)^4 - \left(\chi_{\k}^- \right)^4}{[(i\w_n + i\W_m)^2 - E_{\k}^2][(i\w_n)^2 - E_{\k}^2]} \, .
\end{align}
Note that $\Pi^{ab}(i\W_m) = 0$ by virtue of the considerations mentioned above. Performing the Matsubara sums and using the mean field equation (Eq. 3 in the main text), we find
\begin{align}
    \Pi^{aa}(i\W_m) &= -\left( \frac{\eta_1^2}{g_1} + \frac{\eta_2^2}{g_2} \right) + \int_{\k} \frac{\tanh(E_{\k}/2T)}{2E_{\k}} \; \frac{|\chi_{\k}^+|^2 (i\W_m)^2 - 2\Delta_0^2 |\chi^+_{\k}|^4 -2\Delta_0^2 \left( \chi_{\k}^+\right)^4  }{(i\W_m)^2 - (2E_{\k})^2} \\[5pt]
    \Pi^{bb}(i\W_m) &= -\left( \frac{\eta_1^2}{g_1} + \frac{\eta_2^2}{g_2} \right) + \int_{\k} \frac{\tanh(E_{\k}/2T)}{2E_{\k}} \; \frac{|\chi_{\k}^+|^2 (i\W_m)^2 - 2\Delta_0^2 |\chi^+_{\k}|^4 +2\Delta_0^2  \left( \chi_{\k}^+\right)^4 }{(i\W_m)^2 - (2E_{\k})^2}  \, .
\end{align}
Combining with the Hubbard-Stratonovich terms, we define the inverse propagators $\Df_X^{-1}(i\W_m) = -(\eta_1^2/g_1 + \eta_2^2/g_2) - \Pi^{XX}(i\W_m)$,
\begin{align}
    \Df_a^{-1}(i\W_m) = \int_{\k} \frac{\tanh(E_{\k}/2T)}{2E_{\k}} \; \frac{2\Delta_0^2 |\chi^+_{\k}|^4 + 2\Delta_0^2 \left( \chi_{\k}^+\right)^4 - |\chi_{\k}^+|^2 (i\W_m)^2 }{(i\W_m)^2 - (2E_{\k})^2} \\[5pt]
     \Df_b^{-1}(i\W_m) = \int_{\k} \frac{\tanh(E_{\k}/2T)}{2E_{\k}} \; \frac{2\Delta_0^2 |\chi^+_{\k}|^4 - 2\Delta_0^2 \left( \chi_{\k}^+\right)^4 - |\chi_{\k}^+|^2 (i\W_m)^2 }{(i\W_m)^2 - (2E_{\k})^2} \, .
\end{align}

\subsection{Linear coupling between generalized clapping modes and gauge field}
Finally, we will consider the terms (\ref{couplingsact}) which linearly couple the different modes of the system. Any such couplings which are non-vanishing will generally lead to hybridization between the two modes, such that the true collective modes of the system (which are independent from one another) are linear combinations of the modes studied so far.

First, we calculate the couplings between the relative amplitude and phase modes $a_q$ and $b_q$ and the ABG/gauge field $V_\mu$. Schematically, the linear couplings in (\ref{couplingsact}) take the form 
\begin{equation}
    \begin{split}
        \tr \left( \mathcal{G} \mathbb{V}^{(1)} \mathcal{G} \mathbb{C}\right)  &= \sum_q \left( \, \Pi^{0a}_q \, V_{-q}^0 a_q + \Pi^{0b}_q \, V_{-q}^0 b_q + \Pi^{ia}_q  V^i_{-q} a_q + \Pi^{ib}_q  V^i_{-q} b_q\right) \, .
    \end{split}
\end{equation}
Writing the clapping mode vertex as $\mathbb{C} = -\Lambda_{\k}^a a_q - \Lambda^b_{\k} b_q$ with $\Lambda_{\k}^a = \chi^-_{\k} \, \tau^+ + \chi^+_{\k} \, \tau^-$ and $\Lambda_{\k}^b = i\left(\chi^-_{\k} \, \tau^+ - \chi^+_{\k} \, \tau^- \right)$, the correlation functions above are defined as
\begin{align}
    \Pi^{0a}_q &= -iT\sum_k \, \tr \Big(\mathcal{G}_{k-q/2} \,\tau_z\, \mathcal{G}_{k+q/2} \, \Lambda^a_{\k} \Big) \\
     \Pi^{0b}_q &= -iT\sum_k \, \tr \Big(\mathcal{G}_{k-q/2} \,\tau_z\, \mathcal{G}_{k+q/2} \, \Lambda^b_{\k} \Big) \\
     \Pi^{ia}_q &= \frac{T}{m}\sum_k \, \tr \Big(\mathcal{G}_{k-q/2} \, \mathcal{G}_{k+q/2} \, \Lambda^a_{\k} \Big)\, k^i \\
       \Pi^{ib}_q &= \frac{T}{m}\sum_k \, \tr \Big(\mathcal{G}_{k-q/2} \, \mathcal{G}_{k+q/2} \, \Lambda^b_{\k} \Big)\, k^i \, .
\end{align}
 After evaluation of the Nambu traces, we find, for $\v{q} = 0$, 
\begin{align*}
    \Pi^{0a}(i\W_m) &=-i\Delta_0 T\sum_k \Bigg\{\left[\left(\chi_{\k}^- \right)^2 - \left(\chi_{\k}^+ \right)^2 \right]i\W_m  + \left[\left(\chi_{\k}^- \right)^2 + \left(\chi_{\k}^+ \right)^2 \right]2\xi_{\k}\Bigg\} \frac{1}{[(i\w_n+i\W_m)^2 - E_{\k}^2][(i\w_n)^2 - E_{\k}^2]} \\[5pt]
    \Pi^{0b}(i\W_m) &= -\Delta_0 T\sum_k \Bigg\{\left[\left(\chi_{\k}^- \right)^2 + \left(\chi_{\k}^+ \right)^2 \right]i\W_m  - \left[\left(\chi_{\k}^- \right)^2 - \left(\chi_{\k}^+ \right)^2 \right]2\xi_{\k} \Bigg\} \frac{1}{[(i\w_n+i\W_m)^2 - E_{\k}^2][(i\w_n)^2 - E_{\k}^2]} \\[5pt]
    \Pi^{ia}(i\W_m) &= -\frac{\Delta_0 T}{m}\sum_k \left[\left(\chi_{\k}^- \right)^2 + \left(\chi_{\k}^+ \right)^2 \right]\big(2i\w_n + i\W_m \big) \frac{k^i}{[(i\w_n+i\W_m)^2 - E_{\k}^2][(i\w_n)^2 - E_{\k}^2]} \\[5pt]
    \Pi^{ib}(i\W_m) &= -\frac{i\Delta_0 T}{m}\sum_k \left[\left(\chi_{\k}^- \right)^2 - \left(\chi_{\k}^+ \right)^2 \right]\big(2i\w_n + i\W_m \big)  \frac{k^i}{[(i\w_n+i\W_m)^2 - E_{\k}^2][(i\w_n)^2 - E_{\k}^2]} \, .
\end{align*}
We note that terms linear in $\xi_{\k}$ vanish on account of particle-hole symmetry, and that couplings proportional to $k^i$ will vanish at $\v{q} = 0$ for any even-parity superconducting state by virtue of inversion symmetry. These considerations imply that $\Pi^{0b}(i\W_m)$ is the only non-vanishing coupling at $\v{q} = 0$, and performing the Matsubara sum yields
\begin{equation}
    \Pi^{0b}(i\W_m) = 2\Delta_0 \int_{\k} \frac{\tanh(E_{\k}/2T)}{E_{\k}} \; \left( \chi_{\k}^+ \right)^2 \; \frac{i\W_m}{(i\W_m)^2 - (2E_{\k})^2} \, .
\end{equation}

\subsection{Linear coupling between generalized clapping and Higgs modes}
We now calculate the last remaining coupling between the relative amplitude and phase and Higgs modes, which we write schematically as
\begin{equation}
    \tr \left(\mathcal{G} \mathbb{C}\mathcal{G}\mathbb{H} \right) = \sum_{q} \Big( \Pi^{ha} h_{-q}a_q + \Pi^{hb} h_{-q} b_q \Big)
\end{equation}
where, again for $\v{q} = 0$, we have
\begin{align}
    \Pi^{ha}(i\W_m) &= T \sum_k \tr \left[ \mathcal{G}_{k+q/2}\Lambda^a_{\k} \mathcal{G}_{k-q/2}(\chi_{\k}^+ \,\tau^+ + \chi^-_{\k} \,\tau^- )\right]\\[5pt]
    &= T\sum_k \, \frac{\left[\left(\chi_{\k}^+ \right)^2 + \left(\chi_{\k}^- \right)^2 \right]\Big((i\w_n)^2 - E_{\k}^2 + 2|\Delta_{\k}|^2 + i\w_n \, i\W_m \Big)}{[(i\w_n+ i\W_m)^2 - E_{\k}^2][(i\w_n)^2 -E_{\k}^2]}  \\[5pt]
    \Pi^{hb}(i\W_m) &= T \sum_k \tr \left[ \mathcal{G}_{k+q/2}\Lambda^b_{\k} \mathcal{G}_{k-q/2}(\chi_{\k}^+ \,\tau^+ + \chi^-_{\k} \,\tau^- )\right]\\[5pt]
    &= T\sum_k \, \frac{\left[\left(\chi_{\k}^- \right)^2 - \left(\chi_{\k}^+ \right)^2 \right]\Big((i\w_n)^2 - E_{\k}^2 + 2|\Delta_{\k}|^2 + i\w_n \, i\W_m \Big)}{[(i\w_n+ i\W_m)^2 - E_{\k}^2][(i\w_n)^2 -E_{\k}^2]} \, .
\end{align}
From this, we see $\Pi^{hb}(i\W_m) = 0$ by Eq. 4 in the main text. Performing the Matsubara sums, we have
\begin{align}
    \Pi^{ha}(i\W_m) &= 2\int_{\k} \frac{\tanh(E_{\k}/2T)}{2E_{\k}} \, \left(\chi_{\k}^+ \right)^2  \left(1 + \frac{(i\W_m)^2 -4|\Delta_{\k}|^2}{(i\W_m)^2 - (2E_{\k})^2} \right)  \, .
\end{align}
Using Eq. 4 in the main text and combining with the Hubbard-Stratonovich term in the action, we define the coupling
\begin{equation}
    \tilde{\Pi}^{ha}(i\W_m) = \left(\frac{\eta_1^2}{g_1} - \frac{\eta_2^2}{g_2} \right) + \Pi^{ha}(i\W_m) =  2\int_{\k} \frac{\tanh(E_{\k}/2T)}{2E_{\k}} \, \left(\chi_{\k}^+ \right)^2  \frac{(i\W_m)^2 -4|\Delta_{\k}|^2}{(i\W_m)^2 - (2E_{\k})^2}  \, .
\end{equation}
\subsection{Collective mode spectrum}
Combining the results of the prior sections, the full quadratic action can be written as (suppressing explicit dependences on $i\W_m)$
\begin{equation}
    \begin{split}
        S &= \sum_q \left[\; \Pi^{00}V_{-q}V_q + n_s^{ij} \,V_{-q}^i V_{q}^j - \Df^{-1}_h \, h_{-q}h_q - \Df^{-1}_a \, a_{-q}a_q - \Df^{-1}_b \, b_{-q}b_q \right] \\
        &+ \sum_q \left[\; \tilde{\Pi}^{ha} \, h_{-q}a_q + \Pi^{0b} V_{-q}^0 b_q \right] \, .
    \end{split}
\end{equation}
We see that this action decouples into two sectors, corresponding to amplitude ($h/a$) and phase ($V^0/b$) fluctuations, with the corresponding actions
\begin{align} \label{finalaction}
S_{\text{ampl}} &= -\sum_q \tworow{h_{-q}}{a_{-q}} \sqm{\Df_h^{-1}}{-\frac{1}{2}\tilde{\Pi}^{ha}}{-\frac{1}{2}\tilde{\Pi}^{ha}}{\Df_b^{-1}} \twovec{h_q}{a_q} \\[5pt]
    S_{\text{phase}} &= -\sum_q \tworow{V_{-q}^0}{b_{-q}} \sqm{-\Pi^{00}}{-\frac{1}{2} \Pi^{0b}}{\frac{1}{2}\Pi^{0b}}{\Df_b^{-1}} \twovec{V_q^0}{b_q}
\end{align}
where we have used that $\tilde{\Pi}^{ha}(-i\W_m) = \tilde{\Pi}^{ha}(i\W_m)$ and $\Pi^{0b}(-i\W_m) = -\Pi^{0b}(i\W_m)$. 
We then identify the four independent collective modes of the system from the eigenvalues of each matrix propagator, yielding 
\begin{align}
    D^{-1}_{A,\pm} &= \frac{\Df^{-1}_h + \Df^{-1}_a}{2} \pm \frac{1}{4} \sqrt{\left(\Df_h^{-1} - \Df_a^{-1} \right)^2 +\left( \tilde{\Pi}^{ha} \right)^2} \\[5pt]
    D^{-1}_{P,\pm} &= \frac{-\Pi^{00} + \Df_b^{-1}}{2} \pm \frac{1}{4} \sqrt{\left(\Pi^{00} + \Df_b^{-1}\right)^2 - \left(\Pi^{0b} \right)^2 } \, .
\end{align}
Analytically continuing to real time, $i\W_n \goes \W + i0$, the spectral functions of the collective modes are given by 
\begin{equation}
    \mathcal{A}_{A,\pm} = -\frac{1}{\pi} \, \text{Im} \; D_{A,\pm} \,,  \qquad  \mathcal{A}_{P,\pm} = -\frac{1}{\pi} \, \text{Im} \; D_{P,\pm} \, .
\end{equation}
    
\subsection{Momentum integrals}
We have expressed all of our results thus far in terms of integrals over an internal momentum $\k$ of the form 
\begin{equation}
    I = \int_{\k}\; \frac{1}{2E_{\k}} \, \frac{1}{\W^2 - (2E_{\k})^2} \; f(\phi_{\k}) \, .
\end{equation}
To perform these integrals, we linearize around the Fermi level, writing $\int_{\k} = \nu \int \df \xi \; \int \frac{\df \phi_{\k}}{2\pi}$ where $\nu$ is the density of states at the Fermi level, and we have assumed the system under study is two-dimensional (the generalization to three dimensions is trivial, but leads to more complex angular integrals). 
The integration over $\xi$ can be performed analytically, according to
\begin{equation}
    I = -\nu \int \frac{\df \phi_{\k}}{2\pi} \; f(\phi_{\k}) \; \frac{\sin^{-1}(\W/2|\Delta_{\k}|)}{\W\sqrt{4|\Delta_{\k}|^2 - \W^2}} \, .
\end{equation}
The remaining angular integrals are generally quite complex for generic choices of the form factor, and must be performed numerically.

\end{widetext}


%

\end{document}